\documentclass[
preprintnumbers,superscriptaddress,nofootinbib,prx]{revtex4-2}

\usepackage{amsfonts,amsmath, amssymb,bm}
\usepackage[utf8]{inputenc}
\usepackage{hyperref}
\usepackage{qcircuit}
\usepackage{enumitem}
\usepackage{graphicx}
\graphicspath{{./figures}}

\newcommand{\no}{\nonumber}
\newcommand{\anc}[1]{| #1 \rangle_{\mathrm{ancilla}}}

\newcommand{\phys}[1]{| #1 \rangle_{\mathrm{phys}}}
\newcommand{\Time}[1]{| #1 \rangle_{\mathrm{time}}}
\newcommand{\sub}[1]{| #1 \rangle_{\mathrm{sub}}}
\newcommand{\species}[1]{| #1 \rangle_{\mathrm{species}}}
\newcommand{\direction}[1]{| #1 \rangle_{\mathrm{direction}}}

\newcommand{\tanc}{| \mathrm{ancilla} \rangle}
\newcommand{\tphys}{| \mathrm{phys} \rangle}
\newcommand{\tx}{| \mathrm{phys;} x \rangle}
\newcommand{\ty}{| \mathrm{phys;}y \rangle}
\newcommand{\tz}{| \mathrm{phys;}z \rangle}
\newcommand{\tvx}{| \mathrm{phys;}v_x \rangle}
\newcommand{\tvy}{| \mathrm{phys;}v_y \rangle}
\newcommand{\tvz}{| \mathrm{phys;}v_z \rangle}

\newcommand{\tsub}{| \mathrm{sub} \rangle}
\newcommand{\tspecies}{| \mathrm{species} \rangle}
\newcommand{\tdirection}{| \mathrm{direction} \rangle}

\usepackage{pgf}
\usepackage{xparse}
\usepackage{tikz}
\usetikzlibrary{arrows,arrows.meta,backgrounds,calc,cd,matrix,math,positioning}

\begin{document}
\date{\today}
\title{Quantum Calculation of Classical Kinetic Equations: A Novel Approach for Numerical Analysis of 6D Boltzmann-Maxwell Equations in Collisionless Plasmas Using Quantum Computing}

\author{H. Higuchi}
\email{higuchi.hayato.007@s.kyushu-u.ac.jp}
\affiliation{{\it Graduate School of Science, Kyushu University, Motooka, Nishi-ku, Fukuoka 819-0395, Japan}}

\author{J. W. Pedersen}
\email{pedersen@hep1.c.u-tokyo.ac.jp}
\affiliation{{\it Graduate School of Arts and Sciences, University of Tokyo,
Komaba, Meguro-ku, Tokyo 153-8902, Japan}}

\author{A. Yoshikawa}
\email{yoshikawa.akimasa.254@m.kyushu-u.ac.jp}
\affiliation{{\it Faculty of Science, Kyushu University, Motooka, Nishi-ku, Fukuoka 819-0395, Japan}}

\begin{abstract}

A novel quantum algorithm for solving the Boltzmann-Maxwell equations of the 6D collisionless plasma is proposed. 
The equation describes the kinetic behavior of plasma particles in electromagnetic fields and is known for the classical first-principles equations in various domains, from space to laboratory plasmas.
We have constructed a quantum algorithm for a future large-scale quantum computer to accelerate its costly computation.
This algorithm consists mainly of two routines: the Boltzmann solver and the Maxwell solver. 
Quantum algorithms undertake these dual procedures, while classical algorithms facilitate their interplay.
Each solver has a similar structure consisting of three steps: Encoding, Propagation, and Integration.
We conducted a preliminary implementation of the quantum algorithm and performed a parallel validation against a comparable classical approach.
IBM Qiskit was used to implement all quantum circuits.
\end{abstract}
\maketitle

\tableofcontents

\section{Introduction}
The space plasma environment, extending from the Sun to the magnetosphere-ionosphere-atmosphere, includes regions of frozen conditions, zones of anomalous resistance caused by electromagnetic turbulence, interconnected regions characterized by weakly ionized gas systems in strong magnetic fields, coupled neutral-atmosphere chemical processes, and pure neutral-atmosphere collision systems.
Owing to their complex interactions, an inclusive understanding and forecasting of the space environment remains an elusive goal, even with the advancements in high-performance instrumentation and in-situ observation of satellites.
Therefore, it is imperative to develop space plasma simulations capable of providing comprehensive insights, ranging from local spatial domains to the global schematic.\par
Historically, the development of space plasma simulations has been constrained by computational time, memory capacity, and data storage limitations, resolving complex phenomena with restricted physics at local space scales.
In light of these constraints, space plasma simulations can be divided into two principal scale hierarchies. One approach endeavors to reproduce Macroscopic phenomena using a coarse approximation, whereas the other aims to recreate Microscopic phenomena derived from first principles.
Examples of the former include magnetohydrodynamics (MHD), while the latter include techniques such as particle-in-cell (PIC) or the Vlasov equation (hereafter referred to as the collisionless Boltzmann equation).
The choice between global simulation and comprehensive simulation of physical processes depends on the required space and time scales.
However, several thematic concerns have emerged that require simulation via coupling between scale hierarchies. For example, we describe the plasma instability of the current sheet and the initiation mechanism of magnetic reconnection.
The importance of kinetic effects resulting from ion-electron dynamics during the onset of magnetic reconnection has been demonstrated \cite{Daughton2003,Moritaka2008}.
To emulate this, a multi-hierarchical simulation with inter-domain coupling of MHD and PIC has been developed, which allows to analyze the influence of macroscopic dynamics on the microscopic physics of magnetic reconnection \cite{Usami2009,Usami2014}.\par
In contrast, the collisionless Boltzmann equation requires advanced numerical computations of the 6D distribution function in both space (3D) and velocity (3D) of the particles, and has traditionally been limited to the analysis of low-dimensional, low-resolution or microscopic phenomena.
Given the susceptibility of direct methods to numerical diffusion, the more accurate electromagnetic Vlasov method has been designed and implemented\cite{Umeda2008,Umeda2009,Minoshima2011,Umeda2012a}. 
The considerable progress in its research has allowed the elucidation of numerous authentic physical phenomena through the use of full electromagnetic Vlasov simulation, notwithstanding certain limitations regarding dimensionality and lattice number, which depend on the availability of computational resources\cite{Umeda2010a,Umeda2010b,Umeda2011,Umeda2012b,Umeda2013,Umeda2014}.
Theoretically, the integration of a collision term into the Boltzmann-Maxwell equations provides a comprehensive representation of the collision effects present in the complex coupled magnetosphere-ionosphere-atmosphere system of the Earth. 

However, the current state of simulation technology is such that the fluid equations incorporating these collision effects have not yet been successfully modeled. The effects resulting from ionospheric collisions affect a variety of facets, ranging from auroras to magnetospheric dynamics (e.g. \cite{yoshikawa2013}), and further lead to the manifestation of complex phenomena (e.g. \cite{PBI2016}). Consequently, the collisionality Boltzmann-Maxwell equations encompass a plethora of significant phenomena within their domain of interest that are relevant to space-earth electromagnetics. In an idealized scenario, the entirety of these phenomena could be computed using the collisional Boltzmann-Maxwell equations, eliminating the need for scaling factorial coupled simulations and the reliance on a variety of assumptions. However, performing high-order numerical computations for the first-principles collisional Boltzmann-Maxwell equation requires the establishment of extremely precise numerical methods, coupled with an enormous computational burden $O(L^6)$ (where $L$ is the number of lattices per spatial degree of freedom), which is currently unattainable even with the computational power of today's supercomputers.\par
In recent years, advances in quantum computing, both software and hardware, have demonstrated numerous advantages of quantum algorithms, such as those represented by \cite{Shor1994}. Following Google's achievement of quantum supremacy in 2019 \cite{Google2019}, the pragmatic implementation of quantum computing in plasma simulation, weather forecasting, fluid simulation, and various fields is attracting interest.
In numerical computation, the first paper on solving linear equations with quantum computer, the so-called the HHL algorithm \cite{Harrow2009}, was published.
Subsequently, a quantum algorithm for linear ordinary differential equations (ODE)\cite{Berry2017} and for partial differential equations(PDE)\cite{Childs2021}, and many for fluid simulations have been reported in recent years \cite{Mezzacapo2015,Budinski2022,Steijl2018,Steijl2019,Steijl2023,Arrazola2019,Cao2013,Wang2020,Gaitan2020,Gaitan2021}.
The employed methodologies vary considerably. Some use quantum computational versions of the lattice gas model \cite{Yepez1998,Yepez2001} or the lattice Boltzmann method \cite{Succi2001}, some use quantum Fourier transforms to solve the Poisson equation, some use HHL algorithms and Hamiltonian simulations and Some combine it with the HHL algorithm and Hamiltonian simulations, others reduce from PDEs to ODEs to solve nonlinear ODEs, and so on.
Among them, the quantum lattice Boltzmann method is constructed by considering the streaming operation as Quantum Walk \cite{Aharonov1993}\cite{Succi2015}.
Similarly, a quantum algorithm for the Dirac equation was proposed \cite{Fillion2017}, using the similarity of a sequence of time-evolving operations to Quantum Walk.
And Todorova et al. developed a quantum algorithm for the collisionless Boltzmann equation that performs discrete real and discrete velocity space propagation by Quantum Walk using a discrete-velocity method \cite{ Todorova2020}.
We consider that this method has an advantage over other quantum differential equation solving methods in that it is easier to introduce first-principles collision terms.

\begin{itemize}
\item Collisionless Boltzmann-Maxwell equations with $\boldsymbol{u}$(:velocity) constant and the electromagnetic field $\boldsymbol{E},\boldsymbol{B}$ under vacuum conditions acting one way:
\end{itemize}
\begin{equation}
    \frac{\partial f}{\partial t} + \boldsymbol{u}_{const}\cdot\frac{\partial f}{\partial  \boldsymbol{x}} + \frac{q}{m}(\boldsymbol{E}+\boldsymbol{u}_{const}\times \boldsymbol{B})\cdot\frac{\partial f}{\partial  \boldsymbol{v}} = 0,\no
\end{equation}
\begin{equation}
  \nabla^2 \boldsymbol{E}-\frac{1}{c^2}\frac{\partial^2 \boldsymbol{E}}{\partial t^2}=0,\no
\end{equation}
\begin{equation}
  \nabla^2 \boldsymbol{B} - \frac{1}{c^2} \frac{\partial^2 \boldsymbol{B}}{\partial t^2}=0.\no
\end{equation}
\par
We developed a quantum algorithm for the 6D Boltzmann-Maxwell equations for collisionless plasmas under the above conditions based on the efficient quantum walk circuit\cite{Douglas2009}. In this process, we calculated the time evolution problem of the 6D distribution function with the addition of velocity space, referring to the quantum algorithms for the the discrete velocity method in the the Boltzmann equation\cite{Todorova2020} and the Macro step in the Navier-Stokes equations\cite{Budinski2022}. Thus, the implementation of the collision term, which is the final goal of our project, is much easier and can be developed step by step. 
Furthermore, according to our quantum algorithm, it is simpler and computationally less expensive to solve all regions with the collisionless Boltzmann-Maxwell equations than with Macro-Micro's hierarchically coupled simulators. The quantum computer's most important advantage, the lattice information in the spatial direction, is parallelized into a single state function by encoding amplitude embedding. The results show that the order of the Quantum Volume as the scale of the quantum circuit is $O\left(N_t\left(\log_2(L)\right)^{2}\right)$, which is an improvement over the order of the computational volume $O\left(N_tL^{6}\right)$ of a similar classical algorithm. \par
In the future, we will develop a quantum algorithm for the collisional Boltzmann-Maxwell equations and apply it to the plasma region from the sun to the Earth's magnetosphere-ionosphere-atmosphere. Thus, this will provide a framework in order to understand and fully predict the space plasma environment. At that time, we expect the device to be used is a future fault-tolerant large-scale quantum computer. This paper develops the first quantum algorithm for this purpose and summarizes the methodology and verification results.\par
This paper is organized as follows: Section \ref{sec:Model} and \ref{sec:Numerical} describe the model of numerical computation, Section \ref{sec:Quantum} describes our Quantum Algorithm of Boltzmann solver, and Section \ref{sec:Comparison} compares and verifies the results of the quantum algorithm with similar classical algorithms. In Section \ref{Discussion}, we discuss current issues and future solutions.
\subsection{Governing equations}\label{sec:Model}
We employ the collisionless plasma Boltzmann and Maxwell equations within an electromagnetic field as governing equations.
Specifically, these equations are given by
\begin{itemize}
  \item The collisionless plasma Boltzmann equation with an electromagnetic field:
  \begin{equation}
    \frac{\partial f}{\partial t} + \boldsymbol{u}_{const}\cdot\frac{\partial f}{\partial  \boldsymbol{x}} + \frac{q}{m}(\boldsymbol{E}+\boldsymbol{u}_{const}\times \boldsymbol{B})\cdot\frac{\partial f}{\partial  \boldsymbol{v}} = 0,\label{eq:boltzmann}
\end{equation}
\item Wave equation for the electric field $\boldsymbol{E}$ in vacuum:
\begin{equation}
  \nabla^2 \boldsymbol{E}-\frac{1}{c^2}\frac{\partial^2 \boldsymbol{E}}{\partial t^2}=0,\label{eq:wavefunc_E}
\end{equation}
\item Wave equation for the magnetic field $\boldsymbol{B}$ in vacuum:
 \begin{equation}
  \nabla^2 \boldsymbol{B} - \frac{1}{c^2} \frac{\partial^2 \boldsymbol{B}}{\partial t^2}=0.\label{eq:wavefunc_B}
\end{equation}
\end{itemize}
Where $ f$ is the distribution function of the plasma particles, $ \boldsymbol{u}$ is the fluid velocity of the plasma, which we assume to be constant, $q/m$ is the charge to mass ratio of the particles and $ \boldsymbol{E}$ and $\boldsymbol{B}$ are the electromagnetic fields. The Maxwell equations can be rewritten in the form of wave equations for the electric and magnetic fields respectively, as above, to implement the quantum algorithms more efficiently.
\subsection{Numerical simulation method}\label{sec:Numerical}

For the execution of nonlinear partial differential equations (\ref{eq:boltzmann},\ref{eq:wavefunc_E},\ref{eq:wavefunc_B}) on quantum computers, these equations require discretization by methods such as the finite difference technique or the finite element method. In the following discourse, the finite difference approach is adopted for the Boltzmann-Maxwell equation, resulting in difference equations that are implementable on quantum circuits.\par
Proceeding with the application of the Forward Time Centered Space(FTCS) scheme, we differentiate the Boltzmann equations for collisionless plasma and derive a discretized representation. The differencing equation for the governing equation (\ref{eq:boltzmann}) is given by
\begin{eqnarray}
    f(x,y,z,v_x,v_y,v_z;t+\Delta t) =&& f - \frac{u_x\Delta t}{2\Delta x}(f_{x+\Delta x} -f_{x-\Delta x}) - \frac{u_y\Delta t}{2\Delta y} (f_{y+\Delta y} -f_{y-\Delta y} ) -\frac{u_z\Delta t}{2\Delta z}(f_{z+\Delta z} - f_{z-\Delta z}) \no\\
    && - \frac{q(\boldsymbol{E}+\boldsymbol{u}_{const}\times \boldsymbol{B})_x\Delta t}{2m\Delta v_x}(f_{v_x+\Delta v_x}- f_{v_x-\Delta v_x} )\no\\
    && -\frac{q(\boldsymbol{E}+\boldsymbol{u}_{const}\times \boldsymbol{B})_y\Delta t}{2m\Delta v_y}( f_{v_y+\Delta v_y}- f_{v_y-\Delta v_y})\no\\ 
    && - \frac{q(\boldsymbol{E}+\boldsymbol{u}_{const}\times \boldsymbol{B})_z\Delta t}{2m\Delta v_z}(f_{v_z+\Delta v_z}-f_{v_z-\Delta v_z}),\label{eq:disc_boltzmann}
\end{eqnarray}
where the value of $f(x,y,z,v_x,v_y,v_z;t)$, namely the distribution function at the reference point $x,y,z,v_x,v_y,v_z$ and time $t$, is simply denoted as $f$, and the same at the point deviating by one unit distance in each direction is denoted with subscripts: (e.g.) $f_{x+\Delta x} := f(x+\Delta x,y,z,v_x,v_y,v_z;t)$.\par
 We simplify the difference Boltzmann equation with the following assumption:
\begin{eqnarray}
  \frac{u_x\Delta t}{2\Delta x} = \frac{u_y\Delta t}{2\Delta y} = \frac{u_z\Delta t}{2\Delta z} = 1.\label{eq:condition_boltzmann}
\end{eqnarray}
\par
Similarly, the difference equations for the electric and magnetic fields are given as
  \begin{eqnarray}
    \boldsymbol{E}(x,y,z;t+\Delta t) =&& \left(2-2\frac{\Delta t^{2}}{\mu_0 \epsilon_0}(\frac{1}{\Delta x^{2}}+\frac{1}{\Delta y^{2}}+\frac{1}{\Delta z^{2}})\right)\boldsymbol{E}-\boldsymbol{E}_{t-\Delta t} \no\\
    &&+\frac{\Delta t^2}{\mu_0 \epsilon_0}\left(\frac{1}{\Delta x^{2}}(\boldsymbol{E}_{x+\Delta x}+ \boldsymbol{E}_{x-\Delta x})+\frac{1}{\Delta y^{2}}(\boldsymbol{E}_{y+\Delta y}+ \boldsymbol{E}_{y-\Delta y})+\frac{1}{\Delta z^{2}}(\boldsymbol{E}_{z+\Delta z}+ \boldsymbol{E}_{z-\Delta z})\right),\label{eq:disc_E}\\
    \boldsymbol{B}(x,y,z;t+\Delta t) =&& \left(2-2\frac{\Delta t^{2}}{\mu_0 \epsilon_0}(\frac{1}{\Delta x^{2}}+\frac{1}{\Delta y^{2}}+\frac{1}{\Delta z^{2}})\right) \boldsymbol{B}-\boldsymbol{B}_{t-\Delta t} \no\\
    &&+\frac{\Delta t^2}{\mu_0 \epsilon_0}\left(\frac{1}{\Delta x^{2}}(\boldsymbol{B}_{x+\Delta x}+ \boldsymbol{B}_{x-\Delta x})+\frac{1}{\Delta y^{2}}(\boldsymbol{B}_{y+\Delta y}+ \boldsymbol{B}_{y-\Delta y})+\frac{1}{\Delta z^{2}}(\boldsymbol{B}_{z+\Delta z}+ \boldsymbol{B}_{z-\Delta z})\right),\label{eq:disc_B}
\end{eqnarray}    
where quantities such as $\boldsymbol{E}$ and $\boldsymbol{B}$ are defined in the same manner as $f$ above. 

Furthermore, for simplicity of notation, we set hereafter as the Lorentz force term as
\begin{eqnarray}
  \boldsymbol{F} := \boldsymbol{u}\times \boldsymbol{B} .
\end{eqnarray}
Also, the speed of light $c$ in equation (\ref{eq:wavefunc_E},\ref{eq:wavefunc_B}) is rewritten here using the permittivity and the permeability ($\epsilon_0$ and $\mu_0$) in the vacuum.
Similar to the Boltzmann equation example, we make the following assumption:
\begin{eqnarray}
  \frac{\Delta t^{2}}{\mu_0 \epsilon_0 \Delta x^{2}} = \frac{\Delta t^{2}}{\mu_0 \epsilon_0 \Delta y^{2}} = \frac{\Delta t^{2}}{\mu_0 \epsilon_0 \Delta z^{2}} = 1 .\label{eq:condition_Maxwell}
\end{eqnarray}
Under the postulates of this manuscript, no velocity is obtained from the first-order velocity moment of the distribution function. Given the use of uniform velocities in both the temporal and spatial domains, the discretized magnetic field equation transforms into the propagation equation of the Lorentz force term.
\par
As a result, we obtain the discretized Botzmann-Maxwell equation to be implemented as follows:
\begin{eqnarray}
    f(x,y,z,v_x,v_y,v_z;t+\Delta t) &=& f - (f_{x+\Delta x} -f_{x-\Delta x}) - (f_{y+\Delta y} -f_{y-\Delta y} ) -(f_{z+\Delta z} - f_{z-\Delta z}) \no\\
    && - \frac{q(\boldsymbol{E}+\boldsymbol{F})_x\Delta t}{2m\Delta v_x}(f_{v_x+\Delta v_x}- f_{v_x-\Delta v_x} ) -\frac{q(\boldsymbol{E}+\boldsymbol{F})_y\Delta t}{2m\Delta v_y}( f_{v_y+\Delta v_y}- f_{v_y-\Delta v_y})\no\\ 
    && - \frac{q(\boldsymbol{E}+\boldsymbol{F})_z\Delta t}{2m\Delta v_z}(f_{v_z+\Delta v_z}-f_{v_z-\Delta v_z}),\label{eq:disc_boltzmann_2}\\
  \boldsymbol{E}(x,y,z;t+\Delta t) &=& -4\boldsymbol{E}-\boldsymbol{E}_{t-\Delta t}+  \boldsymbol{E}_{x+\Delta x} +  \boldsymbol{E}_{x-\Delta x}+  \boldsymbol{E}_{y+\Delta y} +  \boldsymbol{E}_{y-\Delta y} +  \boldsymbol{E}_{z+\Delta z} +  \boldsymbol{E}_{z-\Delta z},\label{eq:disc_E_2}\\
  \boldsymbol{F}(x,y,z;t+\Delta t) &=& -4\boldsymbol{F}-\boldsymbol{F}_{t-\Delta t}+  \boldsymbol{F}_{x+\Delta x} +  \boldsymbol{F}_{x-\Delta x}+  \boldsymbol{F}_{y+\Delta y} +  \boldsymbol{F}_{y-\Delta y} +  \boldsymbol{F}_{z+\Delta z} +  \boldsymbol{F}_{z-\Delta z}.\label{eq:disc_B_2}
\end{eqnarray}

This allows us to evolve the values of $f$ and , ($\boldsymbol{E}$, $\boldsymbol{B}$) independently. We call the quantum routines that perform this evolution the Boltzmann solver and the Maxwell solver, respectively.
For the evolution of $f$ (Boltzmann solver), we need the values of $\boldsymbol{E}$ and $\boldsymbol{F}$ at each time step as they appear in the right-hand side of the equation (\ref{eq:disc_boltzmann_2}), so we use the values obtained by the Maxwell solver.
\section{Quantum Algorithm}\label{sec:Quantum}
In this section, a quantum algorithm based on the discretized Boltzmann-Maxwell equations (\ref{eq:disc_boltzmann},\ref{eq:disc_E},\ref{eq:disc_B}) is constructed and implemented on quantum circuits. This quantum algorithm can be divided into two independent routines: the Boltzmann solver and the Maxwell solver.
They take an initial function of $f$ and $(\boldsymbol{E}, \boldsymbol{B})$ as input, respectively.
Both routines fix time and output physical quantities that evolve in one time step according to difference equations (\ref{eq:disc_E_2},\ref{eq:disc_B_2}).
By iterating this one-step evolution many times, we can obtain the value of a physical quantity that has evolved for an arbitrary time step.\par
\begin{figure}[htbp]
\begin{center}
\includegraphics[scale=1.0]{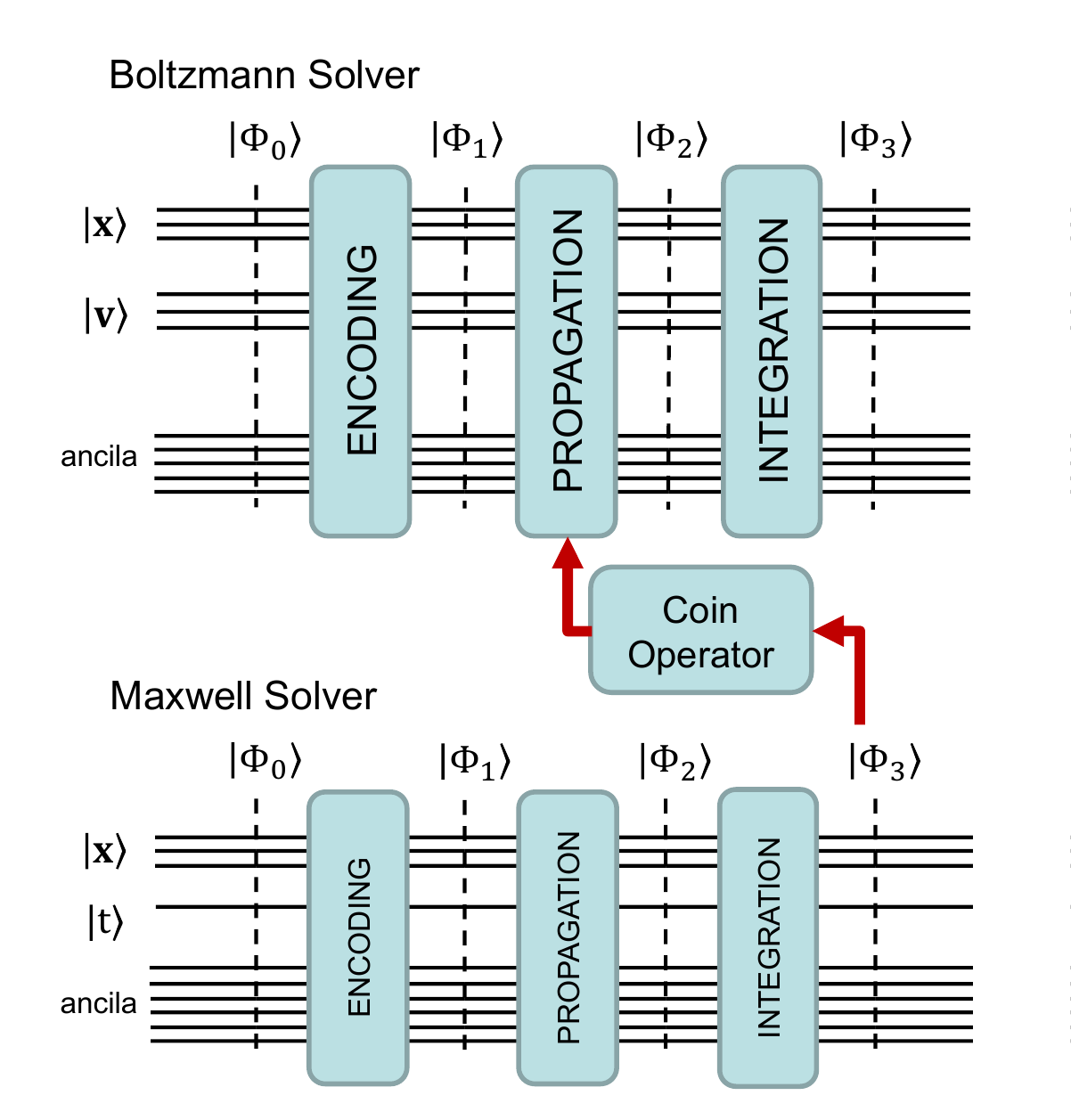}
\caption{A Schematic of the quantum circuit of our algorithm for solving the Boltzmann-Maxwell equations. They consist of two routines that operate on the coin operator.}
\label{fig:QC_Boltzmann-Maxwell}
\end{center}
\end{figure}

The electric and magnetic fields derived by Maxwell solver are incorporated into the Propagation circuit of the Boltzmann solver as shown in the FIG. \ref{fig:QC_Boltzmann-Maxwell}, thereby coupling each routine.
The quantum calculations in this paper are carried out exactly in a way that deals with state vectors using a classical simulator provided by IBM Qiskit.
It is straightforward to construct an authentic quantum algorithm based on measurements. 

\subsection{Boltzmann}
Our Boltzmann solver can be segmented into three principal steps: Encoding, Propagation and Integration.
\subsubsection{Encoding}
First of all, it is necessary to encode the classical information of the physical quantities into the amplitudes of quantum states.
Fixing the number of lattice sites in all spatial and velocity directions to be $L$, $f$ will have $V := L^6$ degrees of freedom. In the encoding step, we associate each of these degrees of freedom with one computational basis and encode the value of $f$ in the amplitude of the corresponding quantum state. Thus, a total of  $V$ bases must be prepared in total, requiring $\lceil \log_2 V \rceil$ qubits. This method of encoding classical information into quantum information amplitudes is commonly referred to as the amplitude embedding technique.\par 
To elucidate the relationship between physical quantities and probability amplitudes, the following conversion from a function $f(\boldsymbol{x},\boldsymbol{v};t)$ to a vector $f_{i} \ , \ (0 \le i \le V-1)$ is implemented.
The subscripts $i$ specify a point in the 6D lattice space.  For example, $i = 0$ corresponds to the origin point $(\boldsymbol{x},\boldsymbol{v}) = (0,0,0,0,0,0)$, and i = 1 represents the value of the distribution function moved by one lattice point in the x direction: $(\boldsymbol{x},\boldsymbol{v}) =  (\Delta x,0,0,0,0,0)$. Namely, the amount of $f_{i}$ follows
\begin{eqnarray}
 \text{(e.g.)} \ \ \ f_{0} &=& f(0,0,0,0,0,0;t=t_r),\\
 f_{1} &=& f(\Delta x,0,0,0,0,0;t=t_r).
\end{eqnarray}
Note that the quantum state does not contain any information about time, since the propagation takes place with fixed time.
We will assume $L = 2^{N_{L}}$ in the following. As evidenced in Section \ref{sec:Comparison}, our actual numerical calculations are executed with $N_{L} = 3 \ (L=8)$.
\par
The first important algorithm in the Encoding step is with a given distribution function at a fixed $t = t_r$ to prepare a quantum state, which we name $\phys{\phi_0}$, with these values in its amplitudes:
\begin{eqnarray}
    \phys{\phi_0} = \sum_{i=0}^{V-1} \tilde{f}_i |i\rangle,\label{eq:phi0}
\end{eqnarray}
where $\tilde{f}$ is the normalized distribution function as follows:
\begin{eqnarray}
  \tilde{f}_{i} = C f_i \ \ , \ \  C = \left(\sum_{i=0}^{V-1} |f_i|^2 \right)^{-1/2}. \label{eq:13}
\end{eqnarray}
At the initial time step of $t = 0$, an arbitrary distribution can be designated as an initial function. Post the second step, the distribution function generated by the Boltzmann solver in the prior step ought to be provided as input. This iterative process allows for the computation of the distribution function at any desired time step. 
This procedure of state preparation can be executed in alignment with Appendix \ref{apendix:unitary}.\par
It should be noted that, within the context of this manuscript, we have formulated the algorithm in a manner that measures $f$ post each step and re-encodes it in the subsequent step, in order to circumvent excessive enlargement of the quantum circuit's depth. This design necessitates $O(V)$ measurements at every time step, failing the advantage of the quantum algorithm. However, it is straightforward to connect each time step seamlessly. Namely, any measurements are required between each time step, implying that such a design will be beneficial when managing large-scale quantum apparatuses in the future. Further discussion on quantum advantage will be given in later sections.\par
The qubits prepared within this context are termed as the physical qubits, denoted as $\tphys$. 
Looking more closely, $\tphys$ is prepared by a total of 6 closed Hilbert spaces corresponding to spatial and velocity degrees of freedom, each having $N_{L} (=\log_2 L)$ qubits. Namely, we write it as
\begin{eqnarray}
    \tphys = \tx\ty\tz\tvx\tvy\tvz
\end{eqnarray}
Subsequent to the Propagation step, the ensuing quantum algorithms necessitate an additional qubit, which depending on their role, is identified as either subnode qubits $\tsub$ or ancilla qubits $\tanc$. As will explaind later the number of subnode and ancilla qubits are fixed to 4 and 1, respectively, regardless of the parameters and physical setup. Thus, the numbers of qubits required by the Boltzmann solver are
\begin{eqnarray}
    N_{\text{phis}} = 6 N_{L} \ , \ N_{\text{sub}} = 4 \ , \ N_{\text{anc}} = 1,
\end{eqnarray}
and the following quantum state is prepared and output in after this Encoding step:
\begin{eqnarray}
    |\phi_1\rangle &=& \phys{\phi_0}\otimes \sub{0} \otimes \anc{0},\\
    &=& \sum_{i = 0}^{V-1} \tilde{f}_i \phys{i} \sub{0} \anc{0}.
\end{eqnarray}

\subsubsection{Propagation}  
In the Propagation step, we partially utilize the tequniques of quantum algorithm method \cite{Douglas2009} and implement an algorithm that multiplies each probability amplitude of $|\phi_1\rangle$ by the coefficient of each term in the discretized equation.
\begin{table}[b]
  \centering
      \caption{ \label{table:table_Boltzmann} 
      The subnode bases and their corresponding physical quantities.
$f, \epsilon,$ and $\sigma$ respectively represent the (unnormalized) distribution function associated with each basis state, the coefficients to be incorporated via the coin operator, and the sign to be multiplied during the integration step. These are the quantities that appear on the right-hand side of the difference equation (\ref{eq:disc_boltzmann_2}).
      }
  \begin{tabular}{c|c|c|c|c}
 \hspace*{1.5mm} $j$\hspace*{1.5mm} &\hspace*{1.5mm} $\sub{j}$\hspace*{1.5mm}& $f_{j}$ & $\epsilon_{j}$& $\sigma_j$ \\ \hline
0 & $| 0000 \rangle$& $f(x,y,z,v_x,v_y,v_z)$& $1$& $+1$\\ 
1  & $| 0001 \rangle$& $f(x+\Delta x,y,z,v_x,v_y,v_z)$& $1$& $-1$\\ 
2  & $| 0010 \rangle$& $f(x-\Delta x,y,z,v_x,v_y,v_z)$& $1$& $+1$\\ 
3  &$| 0011 \rangle$& $f(x,y+\Delta y,z,v_x,v_y,v_z)$& $1$& $-1$\\ 
4  &$| 0100 \rangle$& $f(x,y-\Delta y,z,v_x,v_y,v_z)$& $1$& $+1$\\ 
5  &$| 0101 \rangle$& $f(x,y,z+\Delta z,v_x,v_y,v_z)$& $1$& $-1$\\ 
6  &$| 0110 \rangle$& $f(x,y,z-\Delta z,v_x,v_y,v_z)$& $1$& $+1$\\ 
7  &$| 0111 \rangle$& $f(x,y,z,v_x+\Delta v_x,v_y,v_z)$& $q\frac{E_x(x,y,z)+F_x(x,y,z)\Delta t}{2m\Delta v_x}$& $-1$\\ 
8  &$| 1000 \rangle$& $f(x,y,z,v_x-\Delta v_x,v_y,v_z)$& $q\frac{E_x(x,y,z)+F_x(x,y,z)\Delta t}{2m\Delta v_x}$& $+1$\\ 
9  &$| 1001 \rangle$& $f(x,y,z,v_x,v_y+\Delta v_y,v_z)$& $q\frac{E_y(x,y,z)+F_y(x,y,z)\Delta t}{2m\Delta v_y}$& $-1$ \\ 
10  &$| 1010 \rangle$& $f(x,y,z,v_x,v_y-\Delta v_y,v_z)$& $q\frac{E_y(x,y,z)+F_y(x,y,z)\Delta t}{2m\Delta v_y}$& $+1$\\ 
11  &$| 1011 \rangle$& $f(x,y,z,v_x,v_y,v_z+\Delta v_z)$& $q\frac{E_z(x,y,z)+F_z(x,y,z)\Delta t}{2m\Delta v_z}$& $-1$\\ 
12  &$| 1100 \rangle$& $f(x,y,z,v_x,v_y,v_z-\Delta v_z)$& $q\frac{E_z(x,y,z)+F_z(x,y,z)\Delta t}{2m\Delta v_z}$& $+1$ \\ 
13  &$| 1101 \rangle$& 0 & $0$& $-1$\\ 
14  &$| 1110 \rangle$& 0 & $0$& $+1$\\ 
15  &$| 1111 \rangle$& 0 & $0$& $-1$

  \end{tabular}
 
  \end{table}
To solve the evolution equation (\ref{eq:disc_boltzmann_2}), we need to prepare and add up all the terms that arise in the equation such as
\begin{eqnarray}
    f, \ \mp f_{x\pm\Delta x},\cdots, \mp \frac{q(\boldsymbol{E}+\boldsymbol{F})_x\Delta t}{2m\Delta v_x}f_{v_x\pm \Delta v_x},\cdots.\no
\end{eqnarray}
After passing through the encoding step, we are now in possession of a quantum state $|\phi_1\rangle$, within which the data of the distribution function are encoded in the amplitude. Therefore, by considering an algorithm that multiplies each coefficient such as $\frac{q(\boldsymbol{E}+\boldsymbol{F})_x\Delta t}{2m\Delta v_x}$ by the corresponding state, the amplitudes of all states are updated to the state with the appropriate coefficient appearing in equation (\ref{eq:disc_boltzmann_2}). We will deal with the explicit sign in the equation later. The values of $\boldsymbol{E}$ and $\boldsymbol{F}$ at the certain time step are obtained from Maxwell solver.\par
Subnodes serve to identify the terms that arise at a specific time step, namely $f, f_{x\pm\Delta x},\cdots, f_{v_x\pm\Delta v_x}\cdots$. In total, there are $13 (= 1+ 2\times 6)$ terms: one term $f$, which precedes propagation, and terms propagated by each $\pm 1$ unit for each of the six directions in space and velocity. Hence, $4 (= \lceil 13 \rceil)$ qubits are necessitated as a subnode. It should be noted that this number remains uninfluenced by physical quantities like volume. For simplicity, we have associated them as depicted in TABLE \ref{table:table_Boltzmann}. Here, $\epsilon_{j}$ is the coefficient applied to each term, and $\sigma_j$ is the sign explicitly attributed to each term in TABLE \ref{table:table_Boltzmann}.
In fact, both $\epsilon_{j}$ and $\sigma_j$ are coefficients in the difference equation (\ref{eq:disc_boltzmann_2}), so it is possible to define epsilon to include the sign of $\sigma_j$. However, we choose to distinguish between them because $\epsilon_j$ represents a quantity that depends on a specific assumption (as indicated by the assumption (\ref{eq:condition_boltzmann},\ref{eq:condition_Maxwell}), while $\sigma_j$ is a universally determined quantity. By making this distinction, we think we can minimize the part that we need to be modified based on different assumptions.
\par
As elucidated below, the coin operator is accountable for the multiplication of these coefficients, and the shift operator assumes responsibility for correlating each term with the basis of the subnode.
\par
We can create the appropriate coefficients by first make the subnodes in superpotition using the H-gate. Then apply the diagonal matrix with $\{\epsilon\}$ as components:
\begin{eqnarray}
    \Lambda := \operatorname{diag}(\epsilon_{0}, \epsilon_{1}, \cdots, \epsilon_{15} ).
\end{eqnarray}
The operation with this diagonal matrix is not a unitary and thus it must be embedded in a unitary matrix of larger size. Since the coefficients are real, this procedure can be done easily as explained in the Appendix \ref{apendix:unitary}. Here, we use the ancilla qubit $| a_0 \rangle $ to create a unitary matrix of larger size. We call this whole operator acting on the subnode (and the ancilla qubit) the ``coin operator'' according to the terminology of quantum walk. As a result, we obtain the state after operating the coin operator as follows:
\begin{eqnarray}
  U_{\mathrm{Coin}} |\phi_1\rangle  &=& \sum_{i=0}^{V-1} U_{\mathrm{Coin}} \tilde{f}_i \phys{i} \sub{0} \anc{0},\no\\
  &=& \sum_{i=0}^{V-1} \sum_{j=0}^{15} \tilde{f}_i \tilde{\epsilon}_{j} \phys{i} \sub{j} \anc{0}+ |*\rangle\anc{1},
\end{eqnarray}
where $\tilde{\epsilon}$ represents a normalized quantity. $|*\rangle$ represents the computationally unnecessary states, which are identified by the ancilla qubit being $\anc{1}$.
\par 
Next, so-called increment/decrement gates are applied on both subnode and physical qubits to associate the basis of subnode and physical amount at different points.
The increment/decrement gates are operators that shift one computational basis, respectively. Specifically, those operator satisfy
\begin{eqnarray}
  U_{\mathrm{Incr.}} | i \rangle = | i+1 \rangle,\no\\
  U_{\mathrm{Decr.}} | i \rangle = | i-1 \rangle.\label{eq:19}
\end{eqnarray}
Suppose the periodic boundary condition on the $N$-qubits system:
\begin{eqnarray}
  U_{\mathrm{Incr.}} | 2^{N}-1 \rangle = | 0 \rangle,\no\\
  U_{\mathrm{Decr.}} | 0 \rangle = | 2^{N}-1 \rangle,\label{eq:periodic}
\end{eqnarray}
those operator follow the relation:$U_{\mathrm{Incr.}}^{\dag} = U_{\mathrm{Decr.}}$.
The increment circuit can be specifically configured as follows.

\begin{align}
\raisebox{10mm}{
\raisebox{-9mm}{$ U_{\text{Incr.}}= $\quad}  
  \Qcircuit @C=1em @R=.7em {
  & \ctrl{1} &\ctrl{1}&\ctrl{1} & \gate{X}& \qw\\
  & \ctrl{1} & \ctrl{1} & \targ &\qw& \qw \\
 & \ctrl{1} & \targ & \qw & \qw & \qw\\
 & \targ & \qw & \qw & \qw & \qw
  }  \raisebox{-9mm}{\quad,\quad $ U_{\text{Decr.}}= $\quad}  
  \Qcircuit @C=1em @R=.7em {
  & \ctrlo{1} &\ctrlo{1}&\ctrlo{1} & \gate{X}& \qw\\
  & \ctrlo{1} & \ctrlo{1} & \targ &\qw& \qw \\
 & \ctrlo{1} & \targ & \qw & \qw & \qw\\
 & \targ & \qw & \qw & \qw & \qw 
  }}
\label{eq:fig_shift}
\end{align}

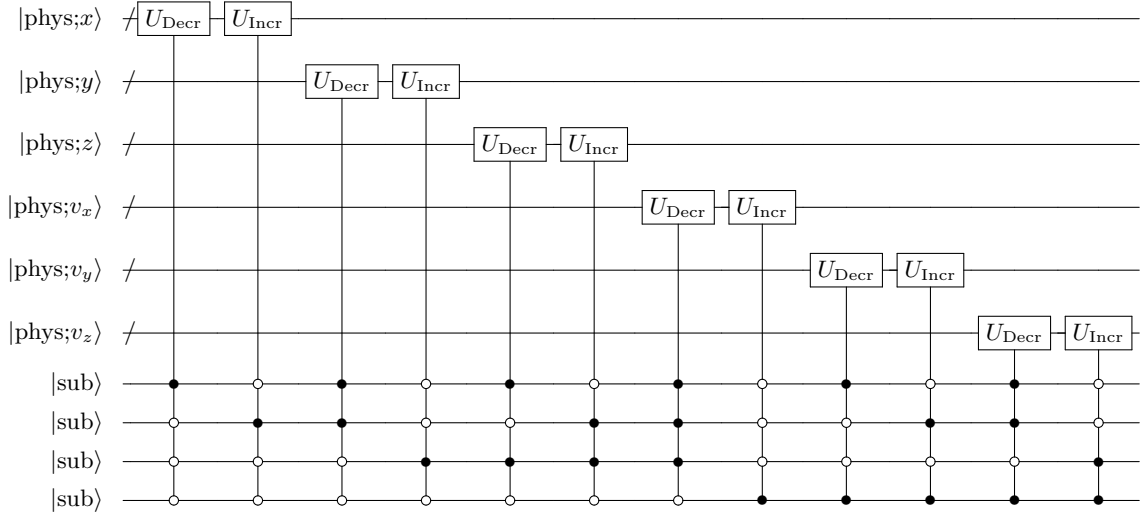
\begin{figure}[htbp]
\begin{center}
\raisebox{0pt}[15pt][195pt]{
\Qcircuit @C=0.3em @R=1.2em {
&\lstick{\tx\hspace{1mm}}& {/} \qw &\gate{U_\text{Decr}}&\qw&\gate{U_\text{Incr}}&\qw&\qw&\qw&\qw&\qw&\qw&\qw &\qw&\qw&\qw&\qw &\qw&\qw&\qw&\qw &\qw&\qw&\qw&\qw &\qw&\qw\no\\
&\lstick{\ty\hspace{1mm}}& {/} \qw &\qw &\qw&\qw&\qw& \gate{U_\text{Decr}} &\qw&\gate{U_\text{Incr}}&\qw&\qw &\qw&\qw&\qw&\qw&\qw &\qw&\qw&\qw&\qw &\qw&\qw&\qw&\qw &\qw&\qw\no\\
&\lstick{\tz\hspace{1mm}}& {/} \qw &\qw&\qw &\qw& \qw &\qw&\qw&\qw& \qw&\gate{U_\text{Decr}} &\qw&\gate{U_\text{Incr}}\qw&\qw&\qw &\qw&\qw&\qw&\qw &\qw&\qw&\qw&\qw &\qw&\qw& \qw\no\\
&\lstick{\tvx\hspace{1mm}}& {/} \qw &\qw&\qw &\qw& \qw &\qw&\qw&\qw& \qw&\qw&\qw &\qw&\qw&\gate{U_\text{Decr}} &\qw&\gate{U_\text{Incr}}\qw&\qw&\qw &\qw& \qw&\qw&\qw &\qw& \qw& \qw \no\\
&\lstick{\tvy\hspace{1mm}}& {/} \qw &\qw&\qw &\qw& \qw &\qw&\qw&\qw& \qw&\qw&\qw &\qw&\qw&\qw&\qw &\qw&\qw&\gate{U_\text{Decr}} &\qw&\gate{U_\text{Incr}}\qw&\qw&\qw&\qw&\qw& \qw \no\\
&\lstick{\tvz\hspace{1mm}}& {/} \qw &\qw&\qw &\qw& \qw &\qw&\qw&\qw& \qw&\qw&\qw &\qw&\qw&\qw&\qw &\qw&\qw&\qw&\qw &\qw&\qw&\gate{U_\text{Decr}} &\qw&\gate{U_\text{Incr}}\qw& \qw \no\\
&\lstick{\tsub\hspace{1mm} }& \qw&\ctrl{-6} &\qw&\ctrlo{-6}&\qw&\ctrl{-5}&\qw&\ctrlo{-5}&\qw&\ctrl{-4}&\qw&\ctrlo{-4}&\qw&\ctrl{-3}&\qw&\ctrlo{-3}&\qw&\ctrl{-2}&\qw&\ctrlo{-2}&\qw&\ctrl{-1}&\qw&\ctrlo{-1}&\qw\no\\
&\lstick{\tsub\hspace{1mm}}& \qw&\ctrlo{-1} &\qw&\ctrl{-1}&\qw&\ctrl{-1}&\qw&\ctrlo{-1}&\qw&\ctrlo{-1}&\qw&\ctrl{-1}&\qw&\ctrl{-1}&\qw&\ctrlo{-1}&\qw&\ctrlo{-1}&\qw&\ctrl{-1}&\qw&\ctrl{-1}&\qw&\ctrlo{-1}&\qw\no\\
&\lstick{\tsub\hspace{1mm}}&\qw&\ctrlo{-1} &\qw&\ctrlo{-1}&\qw&\ctrlo{-1}&\qw&\ctrl{-1}&\qw&\ctrl{-1}&\qw&\ctrl{-1}&\qw&\ctrl{-1}&\qw&\ctrlo{-1}&\qw&\ctrlo{-1}&\qw&\ctrlo{-1}&\qw&\ctrlo{-1}&\qw&\ctrl{-1}&\qw\no\\
&\lstick{\tsub\hspace{1mm}}& \qw&\ctrlo{-1} &\qw&\ctrlo{-1}&\qw&\ctrlo{-1}&\qw&\ctrlo{-1}&\qw&\ctrlo{-1}&\qw&\ctrlo{-1}&\qw&\ctrlo{-1}&\qw&\ctrl{-1}&\qw&\ctrl{-1}&\qw&\ctrl{-1}&\qw&\ctrl{-1}&\qw&\ctrl{-1}&\qw\no
}}
\caption{A Quantum circuit for the shift operators. The increment and decrement operators controlled by subnodes are aligned according to the order of TABLE \ref{table:table_Boltzmann}.}
\label{fig:QC_Shift_Boltzmann}
\end{center}
\end{figure}

 By performing controlled-increment/decrement gates on the subnode as control registers and the physical qubits as target registers, we can map the subnode to a physical quantity on each lattice point. We call this sequential operations as the "shift operator". The circuit of the shift operator is shown in FIG.\ref{fig:QC_Shift_Boltzmann}.

As a result, after applying both the coin operator and the shift operator, we obtain the following state as a final output of this propagation step:

 \begin{eqnarray}
  |\phi_{2}\rangle = \sum_{i=0}^{V -1}\sum_{j=0}^{15} \tilde{\epsilon}_{j}\tilde{f}_{i,j} \phys{i}\sub{j}\anc{0}+ |*\rangle\anc{1}.
\end{eqnarray}
We can articulate the exact correlation between $\tilde{f}i$ and $\tilde{f}{i,j}$ as outlined herein. Initially, we had the capacity to signify the index $i$ as $i = sL + t, \ (0\le s < 6, \ 0\le t < L)$, which, for instance, correlates with the direction $x$ when $s=0$, $y$ when $s=1$, and so forth, and the coordinates of the corresponding directions are symbolized by $t$.
The shift operator moves computational bases in each subspace by $\pm 1$, respecting periodic boundary conditions in each orientation. This $\pm 1$ direction is specified by the index $j$ as shown in TABLE \ref{table:table_Boltzmann}. Therefore, $\tilde{f}_{i,j}$ can be represented as follows:
\begin{eqnarray}
    \tilde{f}_{i,j} = \tilde{f}_{ sL + (t+(-1)^j)\text{mod}L},
\end{eqnarray}
when $i=sL + t, \ (0\le s < 6, \ 0\le t < L)$.
\subsubsection{Integration}
Passing through the encoding and propagation steps so far, we obtain a state in which the all 13 terms arising in the right-hand side of the equation (\ref{eq:disc_boltzmann_2}) for a fixed time step under are encoded in the amplitude of each basis state.
In this step, we perform a superposition of subnode states to compute the sum of all terms and collect them into the amplitude of a single state$\sub{0000}$. However, As a preprocessing step, we need to invert the phases of certain states as explained below.\par
The amplitude of each basis are multiplied by the coefficients in the difference equation \ref{eq:disc_boltzmann_2}, excluding the explicit sign, which is denoted by sigma in TABLE \ref{table:table_Boltzmann}. Therefore, we need to inverse the phase of corresponding state for the terms with a minus sign. This process is also very simple and only requires one application of Z gate as shown in circuit \ref{qc_integration} before applying H gates.\par
Finally, we superimpose all sunode states by applying H as shown in circuit(\ref{qc_integration}).
\begin{align}
\raisebox{20mm}{
\Qcircuit @C=0.8em @R=1.2em {
&\lstick{\tsub\hspace{1mm}}& \qw  &\gate{Z} &\qw&\gate{H}&\qw&\qw\\
&\lstick{\tsub\hspace{1mm}}&  \qw&\qw&\qw &\gate{H}&\qw&\qw\\
&\lstick{\tsub\hspace{1mm}}&  \qw&\qw&\qw &\gate{H}&\qw&\qw\\
&\lstick{\tsub\hspace{1mm}}&  \qw&\qw &\qw&\gate{H}&\qw&\qw
}}\label{qc_integration}
\end{align}
 As a result, the amplitudes of the states from $\sub{0000} $to$ \sub{1111}$ are summed and gathered as the amplitude of $\sub{0000}$ state with equal weighting of 1/4. 
 Therefore, we finally obtain the following state
\begin{eqnarray}
  |\phi_{3}\rangle &=& \frac{1}{4}\sum_{i=0}^{V-1}\sum_{j=0}^{12}\sigma_j\tilde{\epsilon}_{j}\tilde{f}_{i,j}\phys{i}\sub{0000}\anc{0} + |*\rangle\anc{1} .
\end{eqnarray}
With more clear form, we can write
\begin{eqnarray}
    \sum_{j=0}^{12}\sigma_j\tilde{\epsilon}_{j}\tilde{f}_{i,j} &\sim& f - (f_{x+\Delta x} -f_{x-\Delta x}) - (f_{y+\Delta y} -f_{y-\Delta y} ) -(f_{z+\Delta z} - f_{z-\Delta z}) \no\\
    && - \frac{q(\boldsymbol{E}+\boldsymbol{F})_x\Delta t}{2m\Delta v_x}(f_{v_x+\Delta v_x}- f_{v_x-\Delta v_x} ) -\frac{q(\boldsymbol{E}+\boldsymbol{F})_y\Delta t}{2m\Delta v_y}( f_{v_y+\Delta v_y}- f_{v_y-\Delta v_y}),\no\\
    &=& f(x,y,z,v_x,v_y,v_z;t+\Delta t),
\end{eqnarray}
where the distribution function is at the corresponding point of $(x,y,z,v_x, x_y, v_z)$ to the index $i$. Since the normalizing factors of $f$ and $\epsilon$ are involved here, the relation is denoted as ``$\sim$''.\par
According to the resultant state $|\phi_{3}\rangle$, we can measure the physical and subnode qubits and focus on the $\sub{0}$ to obtain a distribution function that is one time step evolved according to the Boltzmann-Maxwell equation. For further time steps, we can use this distribution function as an initial value to input to the first encoding step, and further time evolution can be implemented by performing similar steps.\par
Here are remarks on this algorithm, most of what is touched on here will be discussed more comprehensively in the Section \ref{Discussion}.
First, we asserted that the measurement of the state delivers the value of the distribution function; however, what is specifically attained is the square of the absolute value of the distribution function. Nevertheless, given that the value of the distribution function $f$ is consistently real and non-negative, the precise value of $f$ can be accurately recovered from the measurements. 
On the other hand, $\boldsymbol{E}$ and $\boldsymbol{B}$ handled by Maxwell solver in Appendix \ref{Maxwellsolver} are real but also have negative values, so not exactly the same algorithm can be used.
However, during computation with real quantum algorithms, there isn't a genuine necessity to measure the values of $\boldsymbol{E}$ and $\boldsymbol{B}$. The primary function of the Maxwell solver is simply to convey these values to the Boltzmann solver within the quantum circuit, hence this does not present a significant issue.
If one want to measure $\boldsymbol{E}$ and $\boldsymbol{B}$ values as well, a further ancilla node that identifies the sign must be prepared, and an additional quantum oracle is also needed.\par
Next, Actually measuring $f$ does not lead to quantum advantage. This is because $f$ still has $O(V=L^6)$ degrees of freedom, and it is inevitable to measure it $O(V)$ times in order to obtain full information.
However, this problem can be avoided because what we are physically interested in is not $f$ itself, but the velocity moment quantity obtained by integrating $f$ with respect to velocity $v$.
If we could implement this integral, i.e., just a sum in the discrete system, in an efficient quantum algorithm, the computational complexity would be superior to that of a naive classical algorithm.
Furthermore, we believe that it is possible to reduce the Hilbert space to be measured based on physical conditions such as uniformity with respect to a certain spatial direction, limiting the measurement to the physical space of interest, etc.
\section{Comparison}\label{sec:Comparison}
In this paper, all quantum circuits were exactly simulated by dealing directly with statevectors. Thus it is expected that the results will be in exact agreement with numerical calculations using conventional classical algorithms.
We prepared $L=8$ lattice sites in each spatial and velocity direction and calculated with the volume $V = 8^6$. As for the quantum algorithm $6\times\lceil \log_{2} L\rceil = 18$ qubits were used as $|\text{phys}\rangle$.

And we set $\Delta x=\Delta y=\Delta z=30 \text{m}, \Delta t=10^{-7}\text{s}$, satisfying the assumption (\ref{eq:condition_Maxwell}). Thereby, $v_x=v_y=v_z=3\times 10^{8}\text{m/s}$ is constant at the speed of light. The plasma particles are assumed to be positrons and set $e=1.6\times 10^{-19}\text{C},m_e= 9.1\times 10^{-31}\text{kg}$, so we put $\Delta v_x=\Delta v_y=\Delta v_z=10^{5}\text{m/s}$.
In this section, for simplicity, we re-scale variables $x, y, \cdots$ dividing by the unit $\Delta x, \Delta y, \cdots $ and denote them as coordinates on a lattice space. That is, $x = n$ denotes the point where $x = n \Delta x$ physically.

\subsection{Initial condition}
As the initial distribution function, we employed a simple setup: we set 0 for $(x=1,y=1)$ or $(v_x=1,v_y=1)$, and set 1 for the other spaces. Namely, 
\begin{eqnarray}
\left. f(x,y,z,v_x,v_y,v_z;t=0)\right|_{x=1 \cap y=1} &=& 0, \no\\
\left. f(x,y,z,v_x,v_y,v_z;t=0)\right|_{v_x=1 \cap v_y=1} &=& 0, \no\\
 f(x,y,z,v_x,v_y,v_z;t=0) &=& 1 \  (\text{otherwise}).\no
\end{eqnarray}
This is a simple setup to compare the agreement with the classical algorithm, and in practice it is necessary to give a suitable initial condition corresponding to considering physical phenomena such as plasma.
\begin{figure}[htbp]
\begin{center}
\begin{tabular}{c c}
          
\includegraphics[scale=0.42]{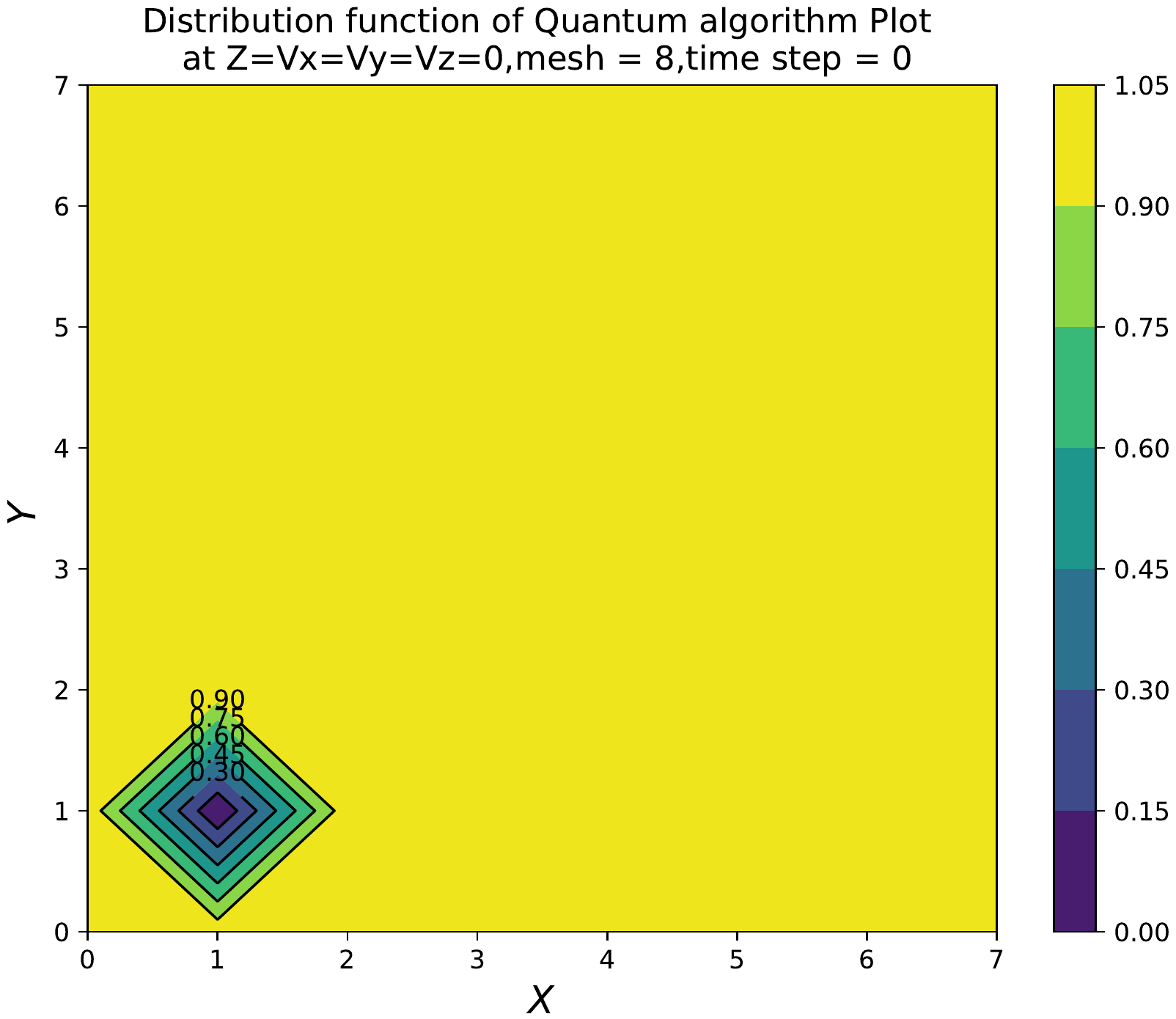}
&
\hspace{3mm}
\includegraphics[scale=0.42]{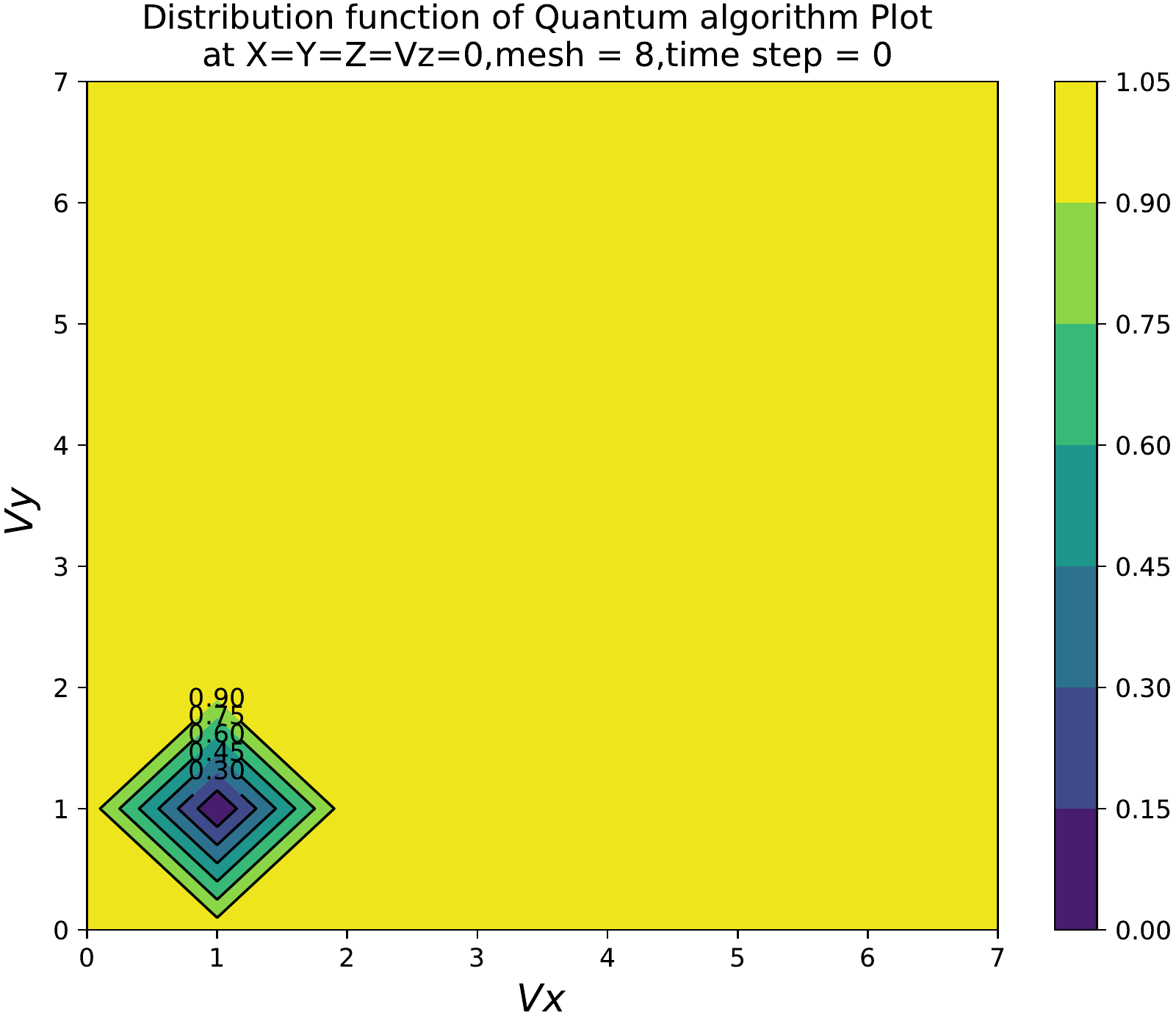}
\\
(a)\hspace*{5mm}&  (b)
\end{tabular}
\caption{ The initial distribution function in the space for (a) the $x-y$ subplane with $z=v_x=v_y=v_z=0$, and (b) the $v_x-v_y$ subplane with  $x=y=z=v_z=0$. This makes it possible to check the influence of electromagnetic fields on propagation in velocity space as well as in real space.}
\label{fig:result}
\end{center}
\end{figure}
\par 
Since we implemented the increment/decrement circuits periodic (\ref{eq:periodic}), the simulation results are also periodic so that the $0$-th and $L$-th lattice points are identical for all directions.
\subsection{Simulation result}
We implemented our quantum algorithm with the input conditions and advanced time evolution from $\text{time step} = 0$ to $\text{time step} = 3$.

\begin{figure}[htbp]
\begin{center}
\begin{tabular}{c c}
          
\includegraphics[scale=0.42]{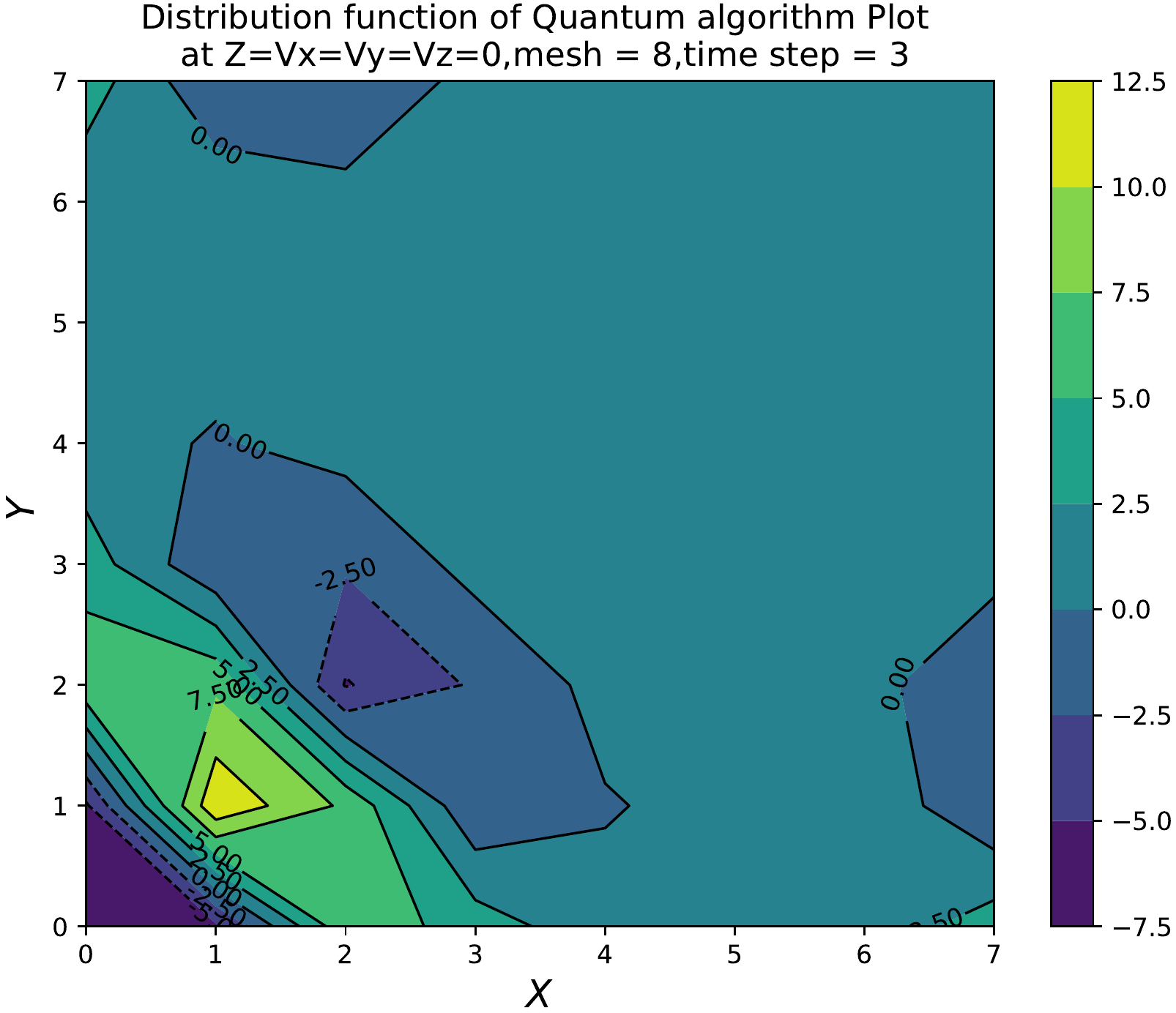}
&
\hspace{3mm}
\includegraphics[scale=0.42]{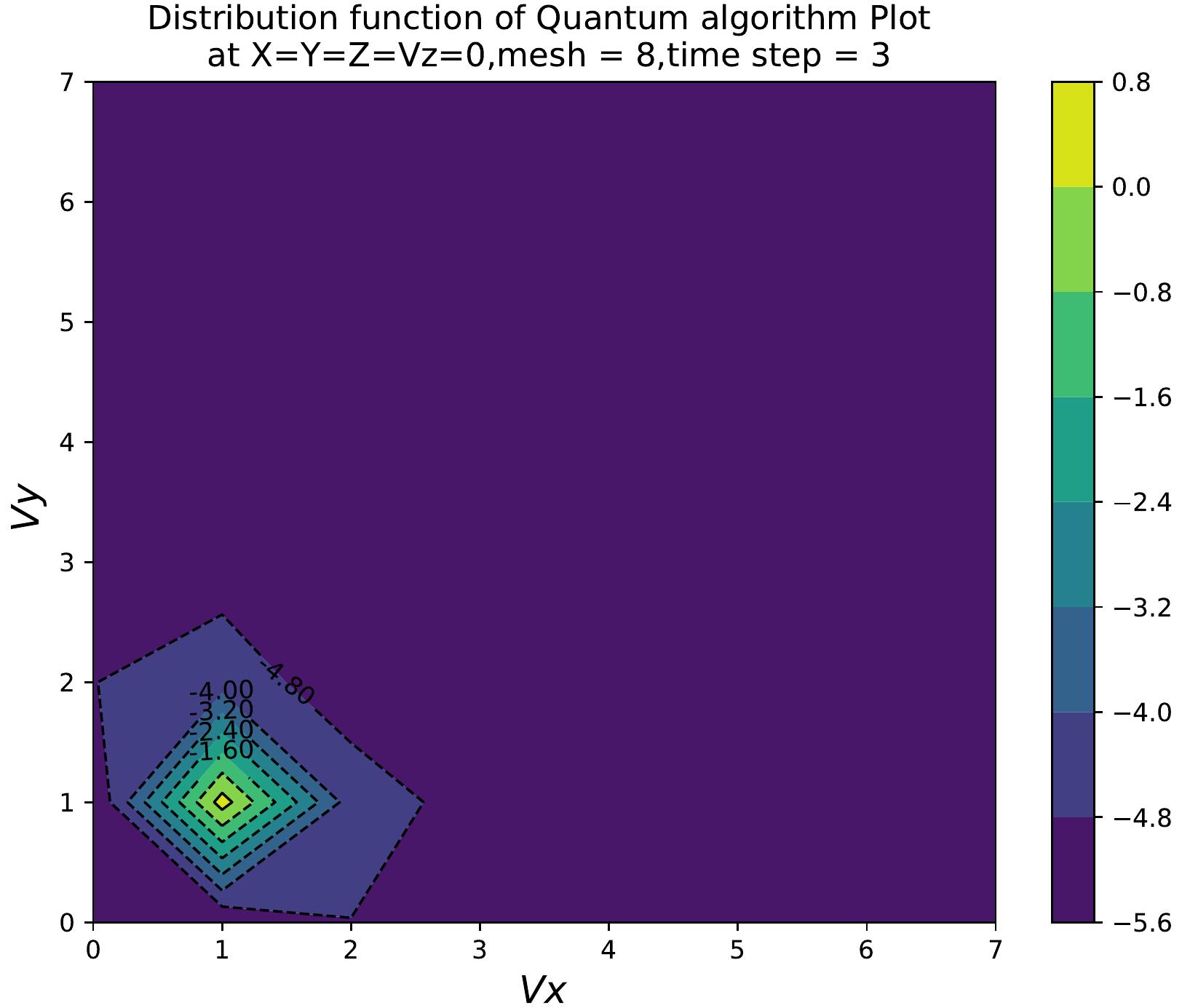}
\\
(a)\hspace*{5mm}& (b)
\end{tabular}
\caption{
Propagated distribution functions after 3 time steps on (a) the $x-y$ subplane with $z=v_x=v_y=v_z=0$ and (b) the $v_x-v_y$ subplane with  $x=y=z=v_z=0$. These results are obtained from the quantum algorithm.
}
\label{fig:result_quantum}
\end{center}
\end{figure}

\begin{figure}[htbp]
\begin{center}
\begin{tabular}{c c}
          
\includegraphics[scale=0.42]{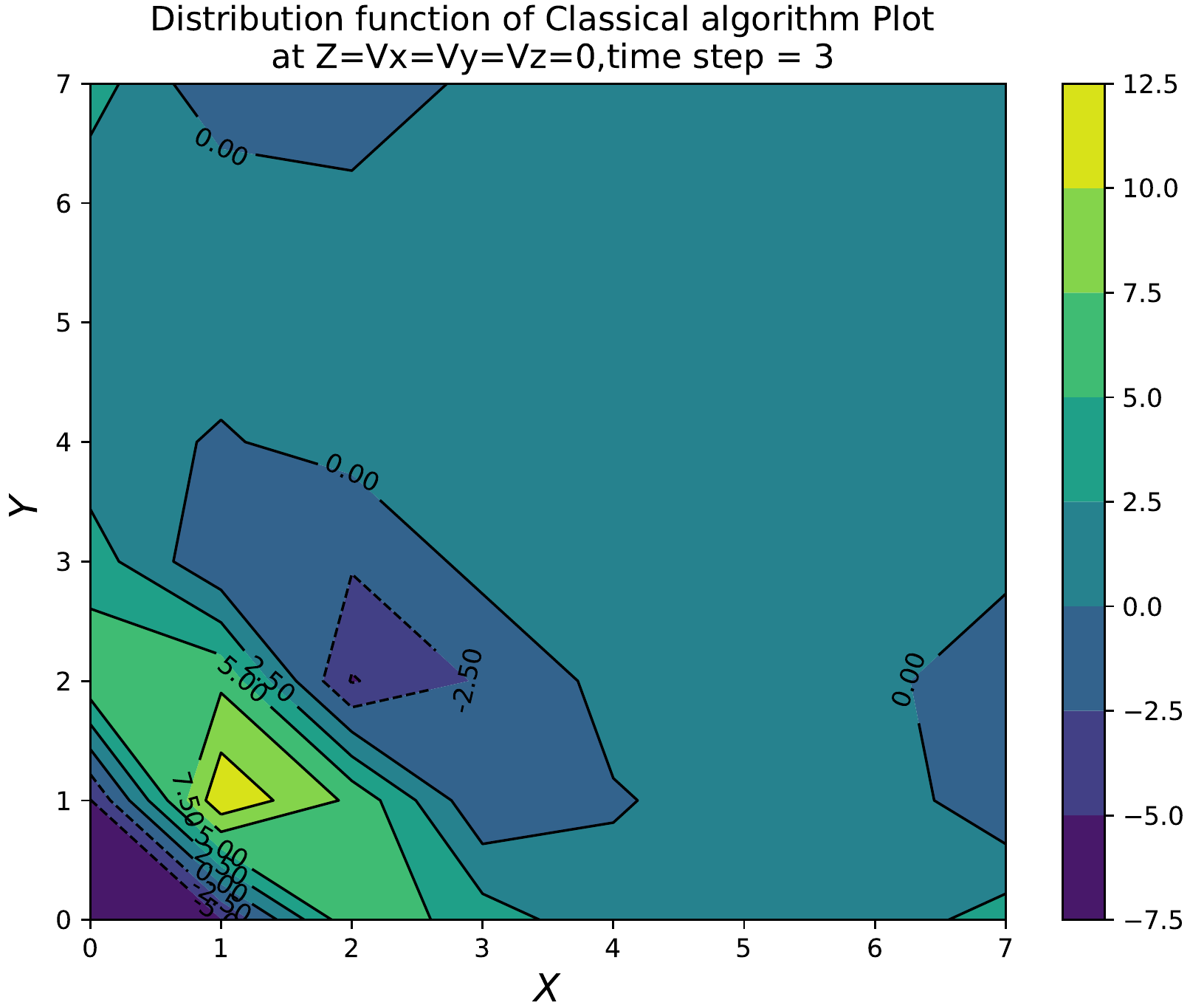}
&
\hspace{3mm}
\includegraphics[scale=0.42]{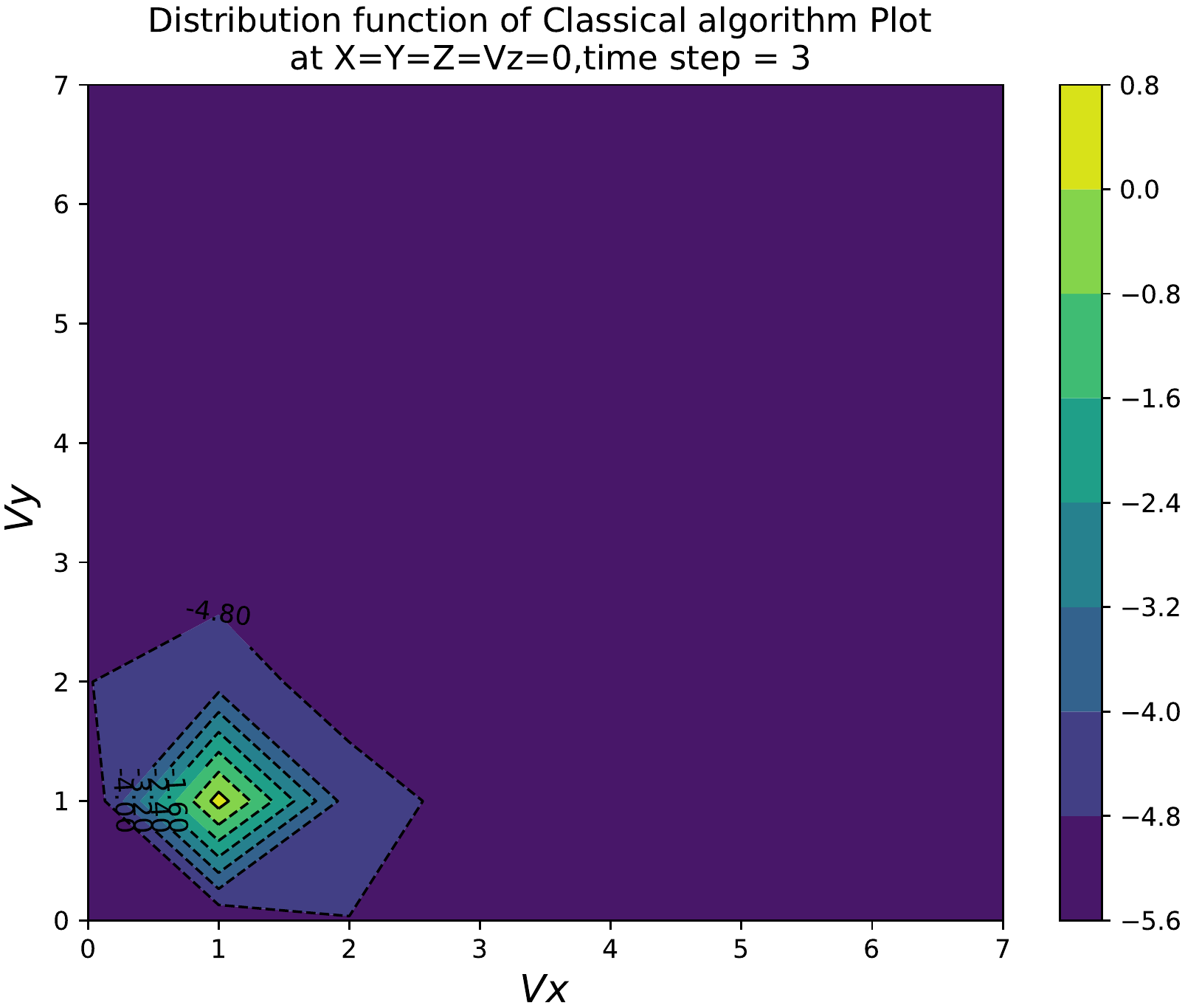}
\\
(a)\hspace*{5mm}& (b)
\end{tabular}
\caption{
The results are based on a classical algorithm of the time evolution of the difference equations (\ref{eq:disc_boltzmann},\ref{eq:disc_E},\ref{eq:disc_B}) using the same FTCS scheme as in this paper, with similar initial and boundary conditions. (a) shows real space propagation at $z=v_x=v_y=v_z=0$ and (b) shows velocity space propagation at $x=y=z=v_z=0$ with time evolution to $\text{time step} = 3$ }
\label{fig:result_classical}
\end{center}
\end{figure}

Comparing FIG. \ref{fig:result_quantum} and FIG. \ref{fig:result_classical}, the simulation results of the quantum algorithm perfectly match those of the classical algorithm with similar conditions and methods. This is because we are simulating exactly with statevector in this case, and the actual results based on measurements will have statistical errors depending on the number of shots.\par
Although $f$ should take values between 0 and 1, this is not the case in FIG. \ref{fig:result_quantum} and FIG. \ref{fig:result_classical}. 
This is a consequence of numerical diffusion due to discretization using the FTCS scheme, which occurs universally in classical algorithms.
As noted in the discussion, the numerical diffusion is reduced by $O(\Delta t)$ in the time direction and $O\left(\left(\Delta x\right)^{2}\right)$ in the space direction , so it is guaranteed to give correct results if the calculation is performed on a sufficiently large system. 

The propagation in real space and velocity space is different, showing that it is acted upon by the electromagnetic field solved with the Maxwell solver. We achieved one of our goals in this paper, that is, the coupling of the Boltzmann equation and the Maxwell equation. However, note that this is a unilateral interaction from the Maxwell equation, since the assumption of uniform velocity and vacuum condition is used.

\section{Discussion}\label{Discussion}
Our plasma simulator is not yet able to cover generic phenomena according to the governing equations (\ref{eq:boltzmann},\ref{eq:wavefunc_E},\ref{eq:wavefunc_B}). This paper is in the middle stage of our project. This means that our plasma simulator does not yet account for velocity inhomogeneity in the convective term of the distribution function, the interaction between electromagnetic fields and plasma particles, and the collisional effects. To add these physical effects, new quantum algorithms must be developed. 
\begin{itemize}
\item Self-consistent collisionless Boltzmann-Maxwell equations interacting with the electromagnetic field by calculating $\rho$ charge density, velocity, and $\boldsymbol{j}$ current density in moment quantities of the distribution function:
\end{itemize}
\begin{equation}
    \frac{\partial f}{\partial t} + \boldsymbol{v}\cdot\frac{\partial f}{\partial  \boldsymbol{x}} + \frac{q}{m}(\boldsymbol{E}+\boldsymbol{v}\times \boldsymbol{B})\cdot\frac{\partial f}{\partial  \boldsymbol{v}} = 0,\no
\end{equation}
\begin{equation}
  \nabla^2 \boldsymbol{E}-\frac{1}{c^2}\frac{\partial^2 \boldsymbol{E}}{\partial t^2}=\frac{1}{\epsilon_0}\nabla\rho+\mu_0\frac{\partial \boldsymbol{j}}{\partial t},\no
\end{equation}
\begin{equation}
  \nabla^2 \boldsymbol{B} - \frac{1}{c^2} \frac{\partial^2 \boldsymbol{B}}{\partial t^2}=-\mu_0\left(\nabla\times \boldsymbol{j}\right).\no
\end{equation}
The next stage will be to improve the current quantum algorithm to the quantum algorithm for the collisionless Boltzmann-Maxwell equation described above. 
To do this, a quantum algorithm that calculates the amount of velocity moments in the distribution function should be developed. Thereby, the electromagnetic field and plasma particles can interact with each other via velocity inhomogeneity, charge density, and current density. This stage can simulate all the complex kinetic effects of collisionless plasma in an electromagnetic field; it simulates macroscopic MHD phenomena that reflect kinetic effects as Micro phenomena. In other words, even macroscopic phenomena can fall back to microscopic phenomena, thus contributing to the complete understanding of the physical process and to the prediction. The domain covers space plasmas in space planetary science, such as the solar surface, and the earth's magnetosphere and astrophysics, such as black hole accretion disks and interstellar winds. 
\begin{itemize}
\item Self-consistent collisional Boltzmann-Maxwell equations interacting with an electromagnetic field, with the addition of a first-principles collision term:
\end{itemize}
\begin{equation}
    \frac{\partial f}{\partial t} + \boldsymbol{v}\cdot\frac{\partial f}{\partial  \boldsymbol{x}} + \frac{q}{m}(\boldsymbol{E}+\boldsymbol{v}\times \boldsymbol{B})\cdot\frac{\partial f}{\partial  \boldsymbol{v}} = Col(f,f^{\prime}),\no
\end{equation}
\begin{equation}
  \nabla^2 \boldsymbol{E}-\frac{1}{c^2}\frac{\partial^2 \boldsymbol{E}}{\partial t^2}=\frac{1}{\epsilon_0}\nabla\rho+\mu_0\frac{\partial \boldsymbol{j}}{\partial t},\no
\end{equation}
\begin{equation}
  \nabla^2 \boldsymbol{B} - \frac{1}{c^2} \frac{\partial^2 \boldsymbol{B}}{\partial t^2}=-\mu_0\left(\nabla\times \boldsymbol{j}\right).\no
\end{equation}
Furthermore, in the final stage, this quantum algorithm will be improved to a quantum algorithm for computing the collision term from the distribution function. By adding a first-principles collision term, the domain of coverage is further extended. It covers the highly complex collisional effects of space plasma versus neutral atmospheres, simulating the ionospheric dynamics of various planetary systems; except for Maxwell solver, it calculates non-equilibrium states of rarefied gases first principles; apply Boltzmann solver and it solves problems of neutrinos and bubble structure in the universe.

We used a finite difference FTCS scheme as our numerical model; the FTCS scheme has numerical errors on the order of $O(\Delta t)$, $O(\Delta x_i^2)$ and $O(\Delta v_{x_i}^2)$ per time evolution. 
Previously, 6D Vlasov simulation research using classical computers has been able to allocate only $L\sim100$ ($L$: lattices per spatial degree of freedom), even using supercomputers.
Therefore, simple numerical methods such as the FTCS scheme are not very appropriate for classical algorithms because of the large numerical errors.
However, in the case of quantum computation with a large-scale quantum computer in a domain that is impossible with a classical computer, the number of lattices per spatial degree of freedom ($\gg100$ lattices) is a very large quantity, and thus the numerical error is inevitably very small.
For example, we estimate that $L>10^{6}$ is needed to simulate the auroral electron acceleration problem in the magnetosphere-ionosphere.
For that very large $L$, the numerical error from the FTCS scheme is small enough.
Moreover, since $L$ increases exponentially with the line increase in hardware logical qubits, the speed of expansion and growth of the computational domain and the speed of improvement in accuracy become exponential.\par

The greatest advantage of quantum algorithms over classical algorithms is massively parallelization. We estimate the Quantum Volume of our quantum algorithm and describe the quantum advantage of the Boltzmann-Maxwell equation. Simply, we will call  Quantum Volume$=$width(number of qubits)$\times$ depth(number of gates) in our quantum algorithm. 
The width of this quantum algorithm is $6\log_2(L)+6$ where $L$ denotes the number of lattice points in each direction.  Comparing to the classical algorithm $O(L)$ computational complexity of the classical algorithm, the fact that it can be expressed in $\log_2(L)$ qubits is a quantum advantage.
On the other hand, the measured quantum circuits for $L=2$, $L=4$, and $L=8$, were found to be approximately $50\times 12\log_2(L)$ per time evolution. In case of time evolution to $\text{Time step} = N_t$, the approximated Quantum Volume would be $3600N_t\log_2(L)\left(\log_2(L)+1\right)$. This is of the order of $O\left(N_t\left(\log_2(L)\right)^{2}\right)$. Compared to the computational volume of a similar classical algorithm $O\left(N_t L^{6}\right)$, the order is improved by compression of 6D spatial information. Thus, the larger $L$ is, the higher the quantum superiority.

Our quantum algorithms are intended for a future large-scale quantum computer, but there remain several issues in terms of efficient algorithms. 
There is a problem of the efficient preparation of the initial distribution function on quantum circuits.
The Encoding step \ref{apendix:unitary} method has the exponential complexity $O\left(2^{N}\right)$ of preparing arbitrary quantum states in a $2^{N}$-dimensional Hilbert space with an $N$ qubit\cite{Zalka1998,Georgescu2014}.
This problem is an important topic in quantum computation, and various efficient methods have been proposed.
For example, Georgescu et al. developed an efficient method to prepare quantum states with polynomial complexity in a number of qubits\cite{Georgescu2014}, and other efficient quantum state initialization methods such as log-concave. Other efficient methods for specific cases, such as log-concave probability distribution functions, have been reported as well\cite{Grover2002}.
Although the initial distribution function varies depending on the physical phenomenon to be simulated, the Maxwell velocity distribution function, for example, is a log-concave probability distribution function and may be efficiently prepared\cite{Todorova2020}.
\par
Our quantum algorithm is more efficient than the classical algorithm in spatial information, but not in the time direction. The reason for this is that the finite difference method of a numerical computation does not allow time information to enter the width of quantum circuits.
The finite difference method is a time-marching-based method for classical numerical calculations using the forward term on the left side of the difference equation.
Due to its nature, one of the degrees of freedom must always be in the depth when implemented in a quantum computer.
Variables that are not set to width are not accelerated, so there are restrictions on the number of lattices with respect to the number of degrees of freedom that can be set to depth, even for large-scale quantum computation.
One simple way to improve this is to rewrite the difference equation of the finite difference method so that the smallest number of lattice degrees of freedom is the evolution parameter instead of time.
Although only one degree of freedom is restricted, this method can keep the depth relatively small.\par
A common problem in quantum differential equation solving is the problem of vanishing time-marching-based measurement probabilities. In general terms, quantum linear system algorithms have an exponentially decreasing measurement probability with respect to the time step, depending on the number of time steps. The quantum algorithm in this study suffers from the same problem.
The first possible solution to this problem is the application of the compression gadget proposed by Fang et al\cite{Fang2023}. This is a time-marching-based quantum differential equation solving method that is independent of time steps by repeating uniform singular value amplification.They verified their implementation on linear ODEs, but it may be applicable to our PDEs.
Next, we also consider the use of different quantum differential equation solving methods as a solution. 
Hamiltonian simulations are a common method for solving quantum differential equations, and the Vlasov-poisson and Vlasov-Maxwell equations have already been used\cite{Toyoizumi2023,Engel2019}. While it is easy to implement the compression gadget \cite{Fang2023} within a Hamiltonian simulation, we consider that it is difficult to implement the nonlinear Boltzmann-Maxwell equations with first-principles collision terms in a Hamiltonian simulation.\par
\section{Summary}
In this paper, a novel quantum algorithm for solving the Boltzmann-Maxwell equation for collisionless plasmas has been formulated; both the Boltzmann and Maxwell equation solvers were structured with a similar quantum circuit. 
To confirm the validity of our quantum algorithm, we performed simulations of the distribution function propagation process under the background electromagnetic field propagation using the Qiskit platform. 
We compared the results of the quantum calculation with the results of the parallel classical calculation and found perfect agreement between them. 
This completes the framework for efficiently solving nonlinear problems in various plasmas, such as space plasmas. 
Prospective endeavors may cultivate the development of a more generalized quantum algorithm for the Boltzmann-Maxwell equation for collisional plasmas, wherein the vacuum condition is eliminated and first-principles collision terms are incorporated.
\section{Acknowledgment}
Discussions during the Yukawa Institute for Theoretical Physics (YITP) summer school YITP-W-22-13 on "A novel numerical approach to quantum field theories” were useful as we started this work. 
HH would like to acknowledge the financial support of the Kyushu University Innovator Fellowship Program (Quantum Science Area).
The work of HH and AY is supported by JSPS KAKENHI Grant Numbers\ JP20H01961 and JP22K21345.
The work of JWP is supported in part by the JSPS Grant-in-Aid for Research Fellow Number\ 22J14732 and the JST SPRING, Grant Number JPMJSP2108.

\appendix
\section{\text{Maxwell solver}}\label{Maxwellsolver}
The basic structure of the Maxwell solver is almost identical to that of the Boltzmann solver. Similar to the Boltzmann solver, the Maxwell solver consists of three steps: encoding, propagation, and integration. The algorithm is briefly described, with special emphasis on the differences to the Boltzmann solver.
\subsection{Encoding}

In Maxwell solver, the physical quantities E and B are written together as g, and develop them simultaneously according to the equations (\ref{eq:disc_E_2},\ref{eq:disc_B_2}).
Since there are no velocity degrees of freedom, only $N_{\text{phys}} = 3\lceil \log_2 L \rceil$ qubit are prepared for $\tphys$, and one additional qubit representing time is also prepared.
$\tsub$ requires $N_{\text{sub}} = 6$ qubit in this case. This is because we need $N_{\text{species}} = 1$ qubit to distinguish the difference of the physical quantity, namely $\boldsymbol{E}$ or $\boldsymbol{B}$, $N_{\text{direction}} =2(=\lceil \log_2 3 \rceil)$ qubits to specify the elements of the vector for them as they are vector, and $N_{\text{term}} = 3(=\lceil \log_2 8 \rceil)$ qubits to indicate 8 terms appearing the equations (\ref{eq:disc_E_2},\ref{eq:disc_B_2}).
Collectively, these are called subnodes, but their roles are actually divided as follows:
\begin{eqnarray}
    \tsub \rightarrow \tspecies \tdirection \tsub.
\end{eqnarray}
These correspondences are shown in Table \ref{table:table_Maxwell} where $\epsilon$ and $\sigma$ represent the the coefficient and explicit sign of each term in the equations (\ref{eq:disc_E_2},\ref{eq:disc_B_2}).
Therefore, using exactly the same algorithm as the Boltzmann solver, we obtain the following state as the outcome of this encoding step:
\begin{eqnarray}
    |\phi_1\rangle = \sum_{i = 0}^{V-1}\sum_{s = 0}^{1}\sum_{d = 0}^{2}\tilde{g}_{i,t,d}\phys{i}\Time{0}\species{s}\direction{d}\sub{0}\anc{0},
\end{eqnarray}
where the subscript i indicates a lattice point using the same rules as in the Boltzmann solver, $g_{i,t,d}$ are given in TABLE \ref{table:table_Maxwell}, and $\tilde{g}$ is normalized $g$. At the first time step we need to specify the initial values for $g$.

\subsection{Propagation}
The structure of the Propagation step in Maxwell solver is fundamentally a Quantum Walk, similar to the Propagation in Boltzmann solver. Thus we need to construct the coin operator and the shift operator. However, the elements of the Coin operator, the time qubits, and the type of subnodes are different. Furthermore, the time increment circuit is used only with respect to the state $\sub{111}$ to use the physical quantity of one previous time. Therefore, in this section, Propagation step generate the states corresponding to the terms propagated in space-time by using the increment and decrement circuits.

The coin operator acts on the subnodes.
\begin{eqnarray}
  U_{\mathrm{coin}}\species{s}\direction{d}\sub{j} = \tilde{\epsilon}_{s,d,j}\species{s}\direction{d}\sub{j},
\end{eqnarray}
where you can also find $\epsilon_{s,d,j}$ in TABLE \ref{table:table_Maxwell} and $\tilde{\epsilon}$ is normalized $\epsilon$. 

\par
One difference from the Boltzmann solver is that the right-hand side of the expression (\ref{eq:disc_E_2},\ref{eq:disc_B_2}) contains a term $g_{i,t-1,s,d}$ that also evolves in the time direction. This effect can be easily implemented by treating time as part of the spatial direction and applying the shift operator in the same way, but note that only the increment circuit is operated since the direction is only negative.
After operating the coin and the shift operator, we obtain the following state as the outcome of this propagation step:
\begin{eqnarray}
  |\phi_{2}\rangle = \sum_{i = 0}^{V-1}\sum_{t = 0}^{1}\sum_{s = 0}^{1}\sum_{d = 0}^{2}\sum_{j = 0}^{7}\tilde{\epsilon}_{s,d,j}\tilde{g}_{i,t,s,d}\phys{i}\Time{t}\species{s}\direction{d}\sub{j}\anc{0}+ |*\rangle\anc{1},
\end{eqnarray}
where $\tilde{g}_{i,t,s,d}$ represents the shift of $\pm 1$ unit in each spatial and the temporal. As for the time direction, $\Time{1}\sub{111}$ and the initial amplitude at $\Time{0}\sub{000}$ are exchanged by the increment circuit (\ref{eq:fig_shift}).The reason for this exchange is because one previous time state is needed to generate a term that propagates in the time direction.

\subsection{Integration}
In contrast to the Boltzmann equation, the Maxwell equation is a second-order differential equation. As a result, the signs $\sigma_j$ that appear in the corresponding difference equation (\ref{eq:disc_boltzmann_2}) differ from those in the Boltzmann equation (as shown in Table \ref{table:table_Maxwell}). In such cases, an controlled-inverse gate, which is shown as follows, should be applied prior to the superposition by the H gate:
\begin{align}
\raisebox{6.5mm}{
  \Qcircuit @C=1em @R=1.4em {
   &\lstick{\tsub\hspace{1mm}}&\ctrl{1}&\qw\\
   &\lstick{\tsub\hspace{1mm}}& \qw & \qw \\
  &\lstick{\tsub\hspace{1mm}}& \ctrl{-1} & \qw 
  } }
\end{align}
\par
The rest of the integration step can use the same method as the Boltzmann solver, but this time we are dealing with different physical quantities, $\boldsymbol{E}$ and $\boldsymbol{B}$, in the same circuit, so we need to sum each of them and not confuse them.
As a result, we can specify the spatial lattice point $(i)$ and the species, and obtain the time-evolved quantities $\boldsymbol{E}, \boldsymbol{B}$ developed in the amplitude of $\sub{000}$.
\begin{table}[htbp]
  \centering

    \caption{
    The subnode bases and their corresponding physical quantities.
$g, \epsilon,$ and $\sigma$ respectively represent the (unnormalized) electromagnetic fields associated with each basis state, the coefficients to be incorporated via the coin operator, and the sign to be multiplied during the integration step. These are the quantities that appear on the right side of the difference equations (\ref{eq:disc_E_2},\ref{eq:disc_B_2}).Here we write only for $\tdirection=\direction{00}$ as an example; $\direction{01}$ and $\direction{10}$ correspond to the $y$- and $z$- components of $\boldsymbol{E}$ and $\boldsymbol{F}$, respectively.
\label{table:table_Maxwell}
    }
  \begin{tabular}{c|c|c|c}
  $\species{s}\direction{d=0}\sub{j}$& $g_{s,d=0,j}$ & $\epsilon_{s,d=0,j}$& $\sigma_j
   $\\\hline
   $\species{0}\direction{00}\sub{000}$& $E_{x}(x,y,z;t)$& $-4$& $+1$\\ 

   $\species{0}\direction{00}\sub{001}$& $E_{x}(x+\Delta x,y,z;t)$& $1$& $+1$\\ 
   $\species{0}\direction{00}\sub{010}$& $E_{x}(x-\Delta x,y,z;t)$& $1$& $+1$\\ 
   $\species{0}\direction{00}\sub{011}$& $E_{x}(x,y+\Delta y,z;t)$& $1$& $+1$\\ 
   $\species{0}\direction{00}\sub{100}$& $E_{x}(x,y-\Delta y,z;t)$& $1$& $+1$\\ 
   $\species{0}\direction{00}\sub{101}$& $E_{x}(x,y,z+\Delta z;t)$& $1$& $+1$\\ 
   $\species{0}\direction{00}\sub{110}$& $E_{x}(x,y,z-\Delta z;t)$& $1$& $+1$\\

    $\species{0}\direction{00}\sub{111}$& $E_{x}(x,y,z;t-\Delta t)$& $1$& $-1$\\
    
   $\species{1}\direction{00}\sub{000}$& $F_{x}(x,y,z;t)$& $-4$& $+1$\\ 

   $\species{1}\direction{00}\sub{001}$& $F_{x}(x+\Delta x,y,z;t)$& $1$& $+1$\\ 
   $\species{1}\direction{00}\sub{010}$& $F_{x}(x-\Delta x,y,z;t)$& $1$& $+1$\\ 
   $\species{1}\direction{00}\sub{011}$& $F_{x}(x,y+\Delta y,z;t)$& $1$& $+1$\\ 
   $\species{1}\direction{00}\sub{100}$& $F_{x}(x,y-\Delta y,z;t)$& $1$& $+1$\\ 
   $\species{1}\direction{00}\sub{101}$& $F_{x}(x,y,z+\Delta z;t)$& $1$& $+1$\\ 
   $\species{1}\direction{00}\sub{110}$& $F_{x}(x,y,z-\Delta z;t)$& $1$& $+1$\\

    $\species{1}\direction{00}\sub{111}$& $F_{x}(x,y,z;t-\Delta t)$& $1$& $-1$
  \end{tabular}
  \end{table}
\section{Construction of our coin operator\label{apendix:unitary}}
In this section we consider an algorithm to multiply a vector to each quantum basis. Let $\Lambda$ denote the multiplying vector:
\begin{eqnarray}
    \Lambda = \left( \lambda_0, \lambda_2, \cdots \lambda_{M-1}\right),
\end{eqnarray}
where we suppose that $\{\lambda\}$ take real values and $\Lambda$ be normalized: $\sum_{i} \lambda_{i}^2 = 1$.\par
To implement this algorithm, we need operate a diagonal matrix $\mathcal{A}$ having entries corresponding to $\Lambda$ but this cannot be done directly because it is not unitary operator in general.
Thus we realized this non-unitary operation by using one ancilla qubit and embedding the matrix $\mathcal{A}$ in a unitary matrix with larger size, which is known as the block encoding method. As $\{ \lambda \}$ are always real, this procedure can easily be implemented as follows:
\begin{eqnarray}
    U = \left(\begin{array}{cc}\mathcal{A} & \mathcal{B} \\ \mathcal{B} & -\mathcal{A}\end{array}\right),
\end{eqnarray}
with 
\begin{eqnarray}
    \mathcal{A} &=& \operatorname{diag}\left( \lambda_1, \lambda_2, \cdots \right),\\
    \mathcal{B} &=&\operatorname{diag}\left( \sqrt{1 - \lambda_1^2}, \sqrt{1 - \lambda_2^2}, \cdots \right).
\end{eqnarray}
After performing this unitary operation on an arbitrary state:
\begin{eqnarray}
    |\psi \rangle = \sum_{i} \alpha_{i} \phys{i}\anc{0}| i \rangle_{phys} | 0 \rangle_{anc},
\end{eqnarray}
we obtain the following state:
\begin{eqnarray}
    |\psi ^{\prime} \rangle &=& U |\psi\rangle, \\
    &=& \sum_{i} \lambda_{i} \alpha_{i}  \phys{i} \anc{0} + |*\rangle\anc{1},
\end{eqnarray}
which we can distinguish desired/unnecessary states with $\anc{0/1}$.

\bibliographystyle{unsrt}
\bibliography{refs}

\begin{thebibliography}{10}

\bibitem{Daughton2003}
William Daughton.
\newblock Electromagnetic properties of the lower-hybrid drift instability in a
  thin current sheet.
\newblock {\em Physics of Plasmas}, 10(8):3103--3119, 07 2003.

\bibitem{Moritaka2008}
Toseo Moritaka and Ritoku Horiuchi.
\newblock {Roles of ion and electron dynamics in the onset of magnetic
  reconnection due to current sheet instabilities}.
\newblock {\em Physics of Plasmas}, 15(9), 09 2008.
\newblock 092114.

\bibitem{Usami2009}
Shunsuke Usami, Hiroaki Ohtani, Ritoku Horiguchi, and Mitsue Den.
\newblock First demonstration of collisionless driven reconnection in a
  multi-hierarchy simulation.
\newblock {\em Plasma and Fusion Research}, 4:049--049, 2009.

\bibitem{Usami2014}
S~Usami, R~Horiuchi, H~Ohtani, and M~Den.
\newblock Multi-hierarchy simulation of collisionless driven reconnection by
  real-space decomposition.
\newblock {\em Journal of Physics: Conference Series}, 561(1):012021, nov 2014.

\bibitem{Umeda2008}
Umeda Takayuki.
\newblock A conservative and non-oscillatory scheme for vlasov code
  simulations.
\newblock {\em Earth, Planets and Space}, 60(7):773--779, 07 2008.

\bibitem{Umeda2009}
Takayuki Umeda, Kentaro Togano, and Tatsuki Ogino.
\newblock Two-dimensional full-electromagnetic vlasov code with conservative
  scheme and its application to magnetic reconnection.
\newblock {\em Computer Physics Communications}, 180(3):365--374, 2009.

\bibitem{Minoshima2011}
Takashi Minoshima, Yosuke Matsumoto, and Takanobu Amano.
\newblock Multi-moment advection scheme for vlasov simulations.
\newblock {\em Journal of Computational Physics}, 230(17):6800--6823, 2011.

\bibitem{Umeda2012a}
Takayuki Umeda, Yasuhiro Nariyuki, and Daichi Kariya.
\newblock A non-oscillatory and conservative semi-lagrangian scheme with
  fourth-degree polynomial interpolation for solving the vlasov equation.
\newblock {\em Computer Physics Communications}, 183(5):1094--1100, 2012.

\bibitem{Umeda2010a}
Takayuki Umeda, Jun-ichiro Miwa, Yosuke Matsumoto, Takuma K.~M. Nakamura,
  Kentaro Togano, Keiichiro Fukazawa, and Iku Shinohara.
\newblock {Full electromagnetic Vlasov code simulation of the
  Kelvin–Helmholtz instability}.
\newblock {\em Physics of Plasmas}, 17(5), 05 2010.
\newblock 052311.

\bibitem{Umeda2010b}
Takayuki Umeda, Kentaro Togano, and Tatsuki Ogino.
\newblock {Structures of diffusion regions in collisionless magnetic
  reconnection}.
\newblock {\em Physics of Plasmas}, 17(5), 05 2010.
\newblock 052103.

\bibitem{Umeda2011}
Takayuki Umeda, Tetsuya Kimura, Kentaro Togano, Keiichiro Fukazawa, Yosuke
  Matsumoto, Takahiro Miyoshi, Naoki Terada, Takuma K.~M. Nakamura, and Tatsuki
  Ogino.
\newblock {Vlasov simulation of the interaction between the solar wind and a
  dielectric body}.
\newblock {\em Physics of Plasmas}, 18(1), 01 2011.
\newblock 012908.

\bibitem{Umeda2012b}
Umeda Takayuki.
\newblock Effect of ion cyclotron motion on the structure of wakes: A vlasov
  simulation.
\newblock {\em Earth, Planets and Space}, 64(2):231--236, 02 2012.

\bibitem{Umeda2013}
Takayuki Umeda, Yosuke Ito, and Keiichiro Fukazawa.
\newblock Global vlasov simulation on magnetospheres of astronomical objects.
\newblock {\em Journal of Physics: Conference Series}, 454(1):012005, aug 2013.

\bibitem{Umeda2014}
Takayuki Umeda, Satoshi Ueno, and Takuma K~M Nakamura.
\newblock Ion kinetic effects on nonlinear processes of the kelvin--helmholtz
  instability.
\newblock {\em Plasma Physics and Controlled Fusion}, 56(7):075006, may 2014.

\bibitem{yoshikawa2013}
A.~Yoshikawa, O.~Amm, H.~Vanhamäki, and R.~Fujii.
\newblock Illustration of cowling channel coupling to the shear alfven wave.
\newblock {\em Journal of Geophysical Research: Space Physics},
  118(10):6405--6415, 2013.

\bibitem{PBI2016}
S.~Ohtani and A.~Yoshikawa.
\newblock The initiation of the poleward boundary intensification of auroral
  emission by fast polar cap flows: A new interpretation based on ionospheric
  polarization.
\newblock {\em Journal of Geophysical Research: Space Physics},
  121(11):10,910--10,928, 2016.

\bibitem{Shor1994}
P.W. Shor.
\newblock Algorithms for quantum computation: discrete logarithms and
  factoring.
\newblock In {\em Proceedings 35th Annual Symposium on Foundations of Computer
  Science}, pages 124--134, 1994.

\bibitem{Google2019}
Frank Arute, Kunal Arya, Ryan Babbush, Dave Bacon, Joseph Bardin, Rami Barends,
  Rupak Biswas, Sergio Boixo, Fernando Brandao, David Buell, Brian Burkett,
  Yu~Chen, Jimmy Chen, Ben Chiaro, Roberto Collins, William Courtney, Andrew
  Dunsworth, Edward Farhi, Brooks Foxen, Austin Fowler, Craig~Michael Gidney,
  Marissa Giustina, Rob Graff, Keith Guerin, Steve Habegger, Matthew Harrigan,
  Michael Hartmann, Alan Ho, Markus~Rudolf Hoffmann, Trent Huang, Travis
  Humble, Sergei Isakov, Evan Jeffrey, Zhang Jiang, Dvir Kafri, Kostyantyn
  Kechedzhi, Julian Kelly, Paul Klimov, Sergey Knysh, Alexander Korotkov, Fedor
  Kostritsa, Dave Landhuis, Mike Lindmark, Erik Lucero, Dmitry Lyakh, Salvatore
  Mandrà, Jarrod~Ryan McClean, Matthew McEwen, Anthony Megrant, Xiao Mi,
  Kristel Michielsen, Masoud Mohseni, Josh Mutus, Ofer Naaman, Matthew Neeley,
  Charles Neill, Murphy~Yuezhen Niu, Eric Ostby, Andre Petukhov, John Platt,
  Chris Quintana, Eleanor~G. Rieffel, Pedram Roushan, Nicholas Rubin, Daniel
  Sank, Kevin~J. Satzinger, Vadim Smelyanskiy, Kevin~Jeffery Sung, Matt
  Trevithick, Amit Vainsencher, Benjamin Villalonga, Ted White, Z.~Jamie Yao,
  Ping Yeh, Adam Zalcman, Hartmut Neven, and John Martinis.
\newblock Quantum supremacy using a programmable superconducting processor.
\newblock {\em Nature}, 574:505–510, 2019.

\bibitem{Harrow2009}
Aram~W. Harrow, Avinatan Hassidim, and Seth Lloyd.
\newblock Quantum algorithm for linear systems of equations.
\newblock {\em Phys. Rev. Lett.}, 103:150502, Oct 2009.

\bibitem{Berry2017}
Dominic~W. Berry, Andrew~M. Childs, Aaron Ostrander, and Guoming Wang.
\newblock Quantum algorithm for linear differential equations with
  exponentially improved dependence on precision.
\newblock {\em Communications in Mathematical Physics}, 356(3):1057--1081, oct
  2017.

\bibitem{Childs2021}
Andrew~M. Childs, Jin-Peng Liu, and Aaron Ostrander.
\newblock High-precision quantum algorithms for partial differential equations.
\newblock {\em Quantum}, 5:574, nov 2021.

\bibitem{Mezzacapo2015}
A.~Mezzacapo, M.~Sanz, L.~Lamata, I.~L. Egusquiza, S.~Succi, and E.~Solano.
\newblock Quantum simulator for transport phenomena in fluid flows.
\newblock {\em Scientific Reports}, 5(1), aug 2015.

\bibitem{Budinski2022}
Budinski Ljubomir.
\newblock Quantum algorithm for the navier–stokes equations by using the
  streamfunction-vorticity formulation and the lattice boltzmann method.
\newblock {\em International Journal of Quantum Information}, 20(02):2150039,
  2022.

\bibitem{Steijl2018}
Ren{\'e} Steijl and George~N. Barakos.
\newblock Parallel evaluation of quantum algorithms for computational fluid
  dynamics.
\newblock {\em Computers and Fluids}, 173:22--28, 2018.

\bibitem{Steijl2019}
Ren{\'e} Steijl.
\newblock Quantum algorithms for fluid simulations.
\newblock In Francisco Bulnes, Vasilios~N. Stavrou, Oleg Morozov, and Anton~V.
  Bourdine, editors, {\em Advances in Quantum Communication and Information},
  chapter~3. IntechOpen, Rijeka, 2019.

\bibitem{Steijl2023}
Ren{\'e} Steijl.
\newblock Quantum circuit implementation of multi-dimensional non-linear
  lattice models.
\newblock {\em Applied Sciences}, 13(1), 2023.

\bibitem{Arrazola2019}
Juan~Miguel Arrazola, Timjan Kalajdzievski, Christian Weedbrook, and Seth
  Lloyd.
\newblock Quantum algorithm for nonhomogeneous linear partial differential
  equations.
\newblock {\em Phys. Rev. A}, 100:032306, Sep 2019.

\bibitem{Cao2013}
Yudong Cao, Anargyros Papageorgiou, Iasonas Petras, Joseph Traub, and Sabre
  Kais.
\newblock Quantum algorithm and circuit design solving the poisson equation.
\newblock {\em New Journal of Physics}, 15(1):013021, jan 2013.

\bibitem{Wang2020}
Shengbin Wang, Zhimin Wang, Wendong Li, Lixin Fan, Zhiqiang Wei, and Yongjian
  Gu.
\newblock Quantum fast poisson solver: the algorithm and complete and modular
  circuit design.
\newblock {\em Quantum Information Processing}, 19(6):170, 2020.

\bibitem{Gaitan2020}
Frank Gaitan.
\newblock Finding flows of a navier–stokes fluid through quantum computing.
\newblock {\em npj Quantum Information}, 6:1--6, 2020.

\bibitem{Gaitan2021}
Frank Gaitan.
\newblock Finding solutions of the navier-stokes equations through quantum
  computing—recent progress, a generalization, and next steps forward.
\newblock {\em Advanced Quantum Technologies}, 4(10):2100055, 2021.

\bibitem{Yepez1998}
Jeffrey Yepez.
\newblock Lattice-gas quantum computation.
\newblock {\em International Journal of Modern Physics C}, 09:1587--1596, 1998.

\bibitem{Yepez2001}
Jeffrey Yepez.
\newblock Quantum lattice-gas model for computational fluid dynamics.
\newblock {\em Phys. Rev. E}, 63:046702, Mar 2001.

\bibitem{Succi2001}
W.~Miller, S.~Succi, and D.~Mansutti.
\newblock Lattice boltzmann model for anisotropic liquid-solid phase
  transition.
\newblock {\em Phys. Rev. Lett.}, 86:3578--3581, Apr 2001.

\bibitem{Aharonov1993}
Y.~Aharonov, L.~Davidovich, and N.~Zagury.
\newblock Quantum random walks.
\newblock {\em Phys. Rev. A}, 48:1687--1690, Aug 1993.

\bibitem{Succi2015}
Sauro Succi, Fran{\c{c}}ois Fillion-Gourdeau, and Silvia Palpacelli.
\newblock Quantum lattice boltzmann is a quantum walk.
\newblock {\em {EPJ} Quantum Technology}, 2(1), may 2015.

\bibitem{Fillion2017}
Fran\ifmmode \mbox{\c{c}}\else~\c{c}\fi{}ois Fillion-Gourdeau, Steve MacLean,
  and Raymond Laflamme.
\newblock Algorithm for the solution of the dirac equation on digital quantum
  computers.
\newblock {\em Phys. Rev. A}, 95:042343, Apr 2017.

\bibitem{Todorova2020}
Blaga~N. Todorova and Ren{\'e} Steijl.
\newblock Quantum algorithm for the collisionless boltzmann equation.
\newblock {\em Journal of Computational Physics}, 409:109347, 2020.

\bibitem{Douglas2009}
B.~L. Douglas and J.~B. Wang.
\newblock Efficient quantum circuit implementation of quantum walks.
\newblock {\em Phys. Rev. A}, 79:052335, May 2009.

\bibitem{Zalka1998}
Christof Zalka.
\newblock Simulating quantum systems on a quantum computer.
\newblock {\em Proceedings of the Royal Society of London. Series A:
  Mathematical, Physical and Engineering Sciences}, 454(1969):313--322, 1998.

\bibitem{Georgescu2014}
I.~M. Georgescu, S.~Ashhab, and Franco Nori.
\newblock Quantum simulation.
\newblock {\em Rev. Mod. Phys.}, 86:153--185, Mar 2014.

\bibitem{Grover2002}
Lov Grover and Terry Rudolph.
\newblock Creating superpositions that correspond to efficiently integrable
  probability distributions, 2002.

\bibitem{Fang2023}
Di~Fang, Lin Lin, and Yu~Tong.
\newblock Time-marching based quantum solvers for time-dependent linear
  differential equations.
\newblock {\em {Quantum}}, 7:955, March 2023.

\bibitem{Toyoizumi2023}
Kiichiro Toyoizumi, Naoki Yamamoto, and Kazuo Hoshino.
\newblock Hamiltonian simulation using quantum singular value transformation:
  complexity analysis and application to the linearized vlasov-poisson
  equation, 2023.

\bibitem{Engel2019}
Alexander Engel, Graeme Smith, and Scott~E. Parker.
\newblock Quantum algorithm for the vlasov equation.
\newblock {\em Phys. Rev. A}, 100:062315, Dec 2019.

\end{thebibliography}
\end{document}